\documentclass[fleqn,usenatbib]{mnras}

\usepackage[T1]{fontenc}
\usepackage{ae,aecompl}

\usepackage{graphicx}	
\usepackage{amsmath}	
\usepackage{amssymb}	

\PassOptionsToPackage{pdfpagelabels=false}{hyperref}

\title[Characterising clusters with GMPhoRCC]{Characterising the optical properties of galaxy clusters with GMPhoRCC}

\author[R. J. Hood and R. G. Mann]{
R. J. Hood,$^{1}$
R. G. Mann,$^{1}$\thanks{E-mail: rgm@roe.ac.uk}
\\
$^{1}$Institute for Astronomy, University of Edinburgh, Royal Observatory, Blackford Hill, Edinburgh, EH9 3HJ, UK
}

\date{Accepted XXX. Received YYY; in original form ZZZ}

\pubyear{2016}

\begin{document}
\label{firstpage}
\pagerange{\pageref{firstpage}--\pageref{lastpage}}
\maketitle

\begin{abstract}

We introduce the Gaussian Mixture full Photometric Red sequence Cluster Characteriser (GMPhoRCC), an algorithm for determining the redshift and richness of a galaxy cluster candidate. By using data from a multi-band sky survey with photometric redshifts, a red sequence colour magnitude relation (CMR) is isolated and modelled and used to characterise the optical properties of the candidate. GMPhoRCC provides significant advantages over existing methods including, treatment of multi-modal distributions, variable width full CMR red sequence, richness extrapolation and quality control in order to algorithmically identify catastrophic failures. We present redshift comparisons for clusters from the GMBCG, NORAS, REFLEX and XCS catalogues, where the GMPhoRCC estimates are in excellent agreement with spectra, showing accurate, unbiased results with low scatter ($\sigma_{\delta z / (1+z)} \sim 0.014$). We conclude with the evaluation of GMPhoRCC performance using empirical Sloan Digital Sky Survey (SDSS) like mock galaxy clusters. GMPhoRCC is shown to produce highly pure characterisations with very low probabilities ($<1\%$) of spurious, clean characterisations. In addition GMPhoRCC is shown to demonstrate high rates of completeness with respect to recovering redshift, richness and correctly identifying the BCG. 

\end{abstract}




\section{Introduction}

Galaxy clusters are excellent probes of cosmology, as the largest observable objects these are great indicators of the large scale structure and evolution of mass distribution in the universe. As this is highly sensitive to the form of the expansion of the universe, their study gives valuable constraints on cosmological models (see \citealp{peebles}, \citealp{sheth}, \citealp{jenkins}, \citealp{maxbcgcosmo}, \citealp{allen} and \citealp{cosmoconst} etc.). In addition clusters provide an excellent opportunity for studying galaxies themselves particularly formation, evolution and the impact of the environment (see \citealp{gladdersredevolution} and \citealp{galev} etc.). With the recent surge in cluster detections, from the Sunyaev-Zel'dovich (SZ) signal in the CMB (\cite{planckclusters}, \cite{spts} etc.), X-ray emission of intracluster medium (ICM) (\cite{xcsxray}, \cite{xclass} etc.), spatial and optical cluster finding (\cite{gmbcg}, \cite{orca}, \cite{redmapper},  etc.), galaxy clusters are proving to be an ever more valuable area of research.

While the most useful cosmological analysis of galaxy clusters involves the study of mass and redshift, these are difficult and time consuming to determine directly requiring gravitational lensing and spectroscopy. Optical characterisation offers quick estimates of cluster properties using multi-band optical photometry alone and with the abundance of the such data from the Sloan Digital Sky Survey (SDSS) \citep{sloan10}, Canada-France-Hawaii Telescope Lensing Survey (CFHTLenS) \citep{cfhtlens}, VLT Survey Telescope (VST) ATLAS \citealp{atlas} and the Panoramic Survey Telescope and Rapid Response System (Pan-STARRS) $3\pi$ survey \citep{panstarrs}, there is significant scope for such analysis.

The main focus of this research is the development of a new characterisation algorithm, the \textbf{G}aussian \textbf{M}ixture full \textbf{Pho}tometric \textbf{R}ed sequence \textbf{C}luster \textbf{C}haracteriser (GMPhoRCC), which aims to provide optical characterisation of potential clusters previously detected by other observations such as X-ray emission. While the specific motivation for this lies with the determination of cluster redshifts for forthcoming XMM Cluster Survey (XCS) (\citealp{romer}, \citealp{xcsoptical} etc.) data releases, GMPhoRCC is designed for general use, providing characterisations for any list of positions of cluster candidates and any multi-colour galaxy catalogue with photometric redshifts.

This paper is structured as follows, Section \ref{sec:characterisation} discusses existing characterisation methods focusing on the motivation and key features desired in a new robust algorithm. Section \ref{sec:pipe} explores the and details of GMPhoRCC with evaluation using comparisons to known and mock clusters following in Section \ref{sec:mocks}, with a detailed investigation of purity, completeness and the effectiveness of the quality control system. Finally this paper concludes with a summary and discussion in Section \ref{sec:concl}. This paper assumes a flat $\Lambda$ CDM cosmology with $\Omega_m = 0.27$, $\Omega_\lambda = 0.73$ and $h=0.71$.

\section{Characterising the optical properties of galaxy clusters}
\label{sec:characterisation}

Many cluster detection/classification algorithms have been developed in recent years, C4 \citep{c4}, maxBCG \citep{maxbcg}, GMBCG \citep{gmbcg} and redMaPPer \citep{redmapper} to name a few. While the exact details vary, the basis of these methods is to isolate the red sequence, early type galaxies with similar metallicities and colours which dominate the members, and use these to infer the bulk properties of the cluster. While optical cluster finders search for additional spatial clustering, the simplest form of red sequence modelling is to find clustering in colour space.

\begin{figure}
  \centering
  \includegraphics[width=240pt]{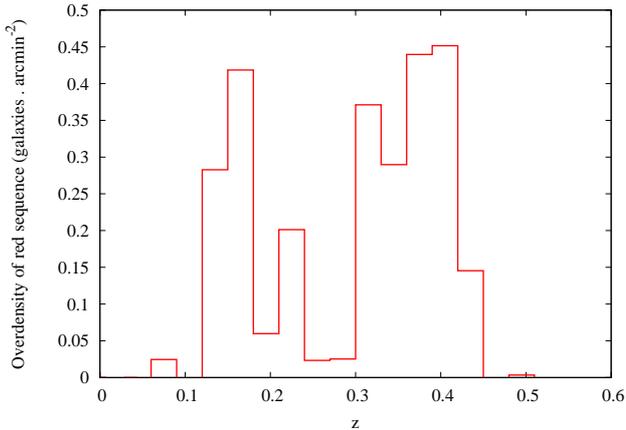} 
  \caption[A histogram of the red sequence overdensity as a function of redshift showing a multi-modal distribution.]{A histogram of the red sequence overdensity as a function of redshift showing a multi-modal distribution resulting from projection effects. These peaks correspond to spatially overlapping clusters at different redshifts and without any additional information it is difficult to determine which represents the target cluster.}
  \label{fig:highbad}
\end{figure}

The simplest case of characterisation, relates to a well defined easily observed red sequence as an overdensity in colour or redshift space, however many clusters do not conform to this due to projection effects and background fluctuations. Figure \ref{fig:highbad} demonstrates the redshift clustering of a field with two overlapping clusters. Due to the projection effect it is unclear which peak represents the target cluster and methods looking for maximum overdensities such as that from \cite{high} may fail to adequately describe the situation. With these multi-modal distributions common, found to be present in $\sim 40$ percent of the GMBCG catalogue, it is clear any stand alone characterisation algorithm must account for these ambiguous cases. 

While a simple colour overdensity is a good approximation, the red sequence itself is described by a colour magnitude relation (CMR). The CMR relates to the physical properties of the cluster, where slope encodes the mass-metallicity relation and scatter age variation etc. (\citealp{gladdersredevolution}), and hence it is desirable to model the full CMR rather than simple overdensities as is the case with GMBCG, XCS etc.

With the red sequence isolated it remains to determine cluster properties. Redshift can be determined using red sequence colour-redshift models as in the cases of \cite{maxbcg} and \cite{xcsoptical}, however this introduces model dependence and additional complexity in the analysis. In the case of GMBCG, redshift is obtained from the photometric estimate of the BCG. Although relying on correct identification, as a bright galaxy the photometric estimate is more easily obtained.

The following lists several initial features drawn from existing algorithms which have driven the development of GMPhoRCC.

\begin{enumerate}
\item Red sequence detection: GMPhoRCC will isolate the red sequence and use these galaxies to infer the optical properties of the cluster.
\item Photometric redshifts: No assumed colour-redshift model.
\item Full red sequence CMR: The red sequence will be described by a full CMR determined by the GMPhoRCC.
\item Multiple red sequence bands: To maximise the efficiency of the GMPhoRCC and to cover a large range of cluster redshifts, multiple redshift dependent colour bands will be necessary as demonstrated in \cite{gmbcg}.
\item Multi-modal distributions: Without resorting to a full finder approach where overlapping clusters can be separately identified and analysed, GMPhoRCC must deal with multi-modal distributions as several potential clusters.
\item Quality Control: Extending beyond simple error analysis it is necessary to assess the probability of catastrophic failure. By introducing quality control, several subsets can be produced where problem clusters and possible outliers can be identified and removed to produce a clean subset. 
\end{enumerate}

\section{GMPhoRCC}
\label{sec:pipe}

\begin{figure}
  \centering
  \includegraphics[width=94pt]{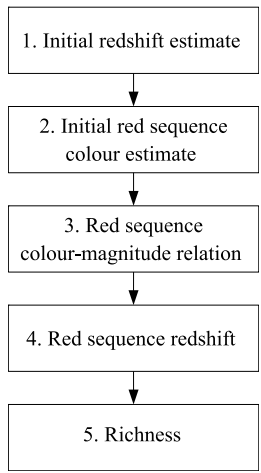} 
  \caption[A flowchart summarising the basic steps of GMPhoRCC to determine the red sequence CMR, cluster redshift and richness.]{A flowchart summarising the basic steps of GMPhoRCC to determine the red sequence CMR, cluster redshift and richness.}
  \label{fig:overviewflow}
\end{figure}

GMPhoRCC is designed as an optical follow-up tool to confirm and characterise galaxy cluster candidates. The main feature of GMPhoRCC is to identify the red sequence and use the properties of these galaxies to analyse the cluster. A basic outline of the procedure used to isolate the red sequence and ultimately determine redshift and richness is shown in Figure \ref{fig:overviewflow}. While each of these steps is explored in detail in subsequent sections it is first noted that many of these require the modelling of cluster distributions such as colour and photometric redshifts which, as shown by \cite{gmbcgerror}, can be approximated by a Gaussian mixture. 

\subsection{Modelling Cluster Distributions with Error-Corrected Gaussian Mixtures}
\label{sec:ecem}

Galaxy distributions, whether colour or redshift, are well modelled by Gaussian mixtures which use a sum of several Gaussian components to describe any features or substructures. Fitting the mixture model proceeds using the error-corrected expectation maximization procedure from \cite{gmbcgerror} which accounts for associated measurement errors. Considering the distribution of galaxy colours as an example, the probability of galaxy $n$ having true colour $\tilde{c}_n$ given the parameters $\theta$ is defined below:
\begin{equation}
\label{eq:gm}
p(\tilde{c}_n \;|\; \theta) = \sum_k{N(\tilde{c}_n \;|\; \mu_k, \sigma_k )P(k)}
\end{equation}
\begin{equation}
N(\tilde{c}_n\;|\;\mu_k, \sigma_k ) = \frac{1}{\sqrt{2\pi\sigma_k^2}} \exp{-\frac{(\tilde{c}_n-\mu_k)^2}{2\sigma_k^2}}
\end{equation}

\noindent where $\theta$ represents the collective parameters of the model namely $\mu_k$, $\sigma_k$, $P(k)$, the component means, standard deviations and weights respectively. Combining these with Gaussian measurement errors for all the galaxies, \cite{gmbcgerror} have shown that this leads to the following form from the likelihood of the parameters given the data:
\begin{equation}
\label{eq:lik}
p(\theta\;|\;c_n)=\prod^{N}_{n=1}\left(\sum^{K}_{k=1} \frac{P(k)}{\sqrt{2\pi(\sigma_k^2+\delta_n^2)}} \exp{\left[-\frac{(c_n-\mu_k)^2}{2(\sigma_k^2+\delta_n^2)} \right]}\right)
\end{equation}

\noindent where $c_n$ is the observed galaxy colour with Gaussian error $\delta_n$, $K$ is the total number of components and $N$ is the number of galaxies. Maximising this likelihood using the expectation maximisation procedure of \cite{gmbcgerror} gives the optimised parameters.

In order to apply this modelling to the distributions of cluster members it is necessary to account for the effect of the background when analysing a candidate field. Rather than modelling the whole field and selecting components to separately describe the cluster and background as explored by the GMBCG algorithm \cite{gmbcg}, a subtraction approach is used.

\begin{equation}
GM_{cluster} = GM_{field} - GM_{background}
\end{equation}

\noindent where $GM$ represents a Gaussian mixture density model. Modelling clusters in this manner with background subtraction helps removes the ambiguity in component selection seen in GMBCG. Values of interest, such as an approximate red sequence colour from the colour distribution, are determined with an associated uncertainty from the width of the distributions around the peaks. While this section describes in detail the Gaussian mixture fitting procedure for a given area, the selection of the cluster area, background and field is explored in Section \ref{sec:pipered}.

Applying this to the colour distribution of cluster GMBCG J197.87292-01.34109 indeed shows the Gaussian mixture to be a good representation of the cluster around a $g-r$ colour of $\sim 1.2$ magnitudes.

\begin{figure}
  \centering
  \includegraphics[width=240pt]{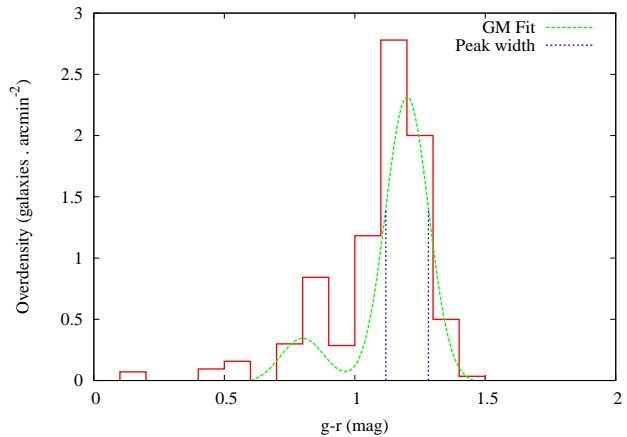} 
  \caption[A histogram of the $g-r$ colour overdensity of GMBCG-1 showing the best fit Gaussian mixture.]{A histogram of the $g-r$ colour overdensity of GMBCG J197.87292-01.34109 showing the best fit Gaussian mixture highlighting the associated width used as an asymmetric error bar around the peak.}
  \label{fig:gmcol}
\end{figure}
	
\subsection{Red sequence CMR and Redshift}
\label{sec:pipered}

With the framework in place to model the various cluster distributions from optical data the first main goal of GMPhoRCC is to identify the red sequence, modelling a full CMR with intrinsic scatter. As the previous section described, GMPhoRCC models cluster distributions with background subtraction, hence the first step is the extraction of the cluster region and background from the candidate field. The cluster region is taken as the cone with a $30$ arc minute radius centred on the detection observation, for example, the peak of the X-ray emission. The local background is taken as the annulus around this cone up to a radius of $60$ arc minutes.

\subsubsection{Initial Redshift Estimate}

To aid the isolation of the red sequence, initial redshift and colour estimates for the cluster are determined which allow the selection of an appropriate colour band and provides broad filtering to remove field galaxies. While this is perhaps best achieved by modelling the joint colour-redshift density distribution, extending the fitting procedure of Section \ref{sec:ecem} to higher dimensions is not trivial. Even without the error correction the fitting procedure often fails to converge, fails to recover fine structure and is highly sensitive to the initial estimate of the parameters. Hence separate, error-corrected modelling of redshift and colour proceeds.

\begin{figure*}
  \centering
  \includegraphics[width=390pt]{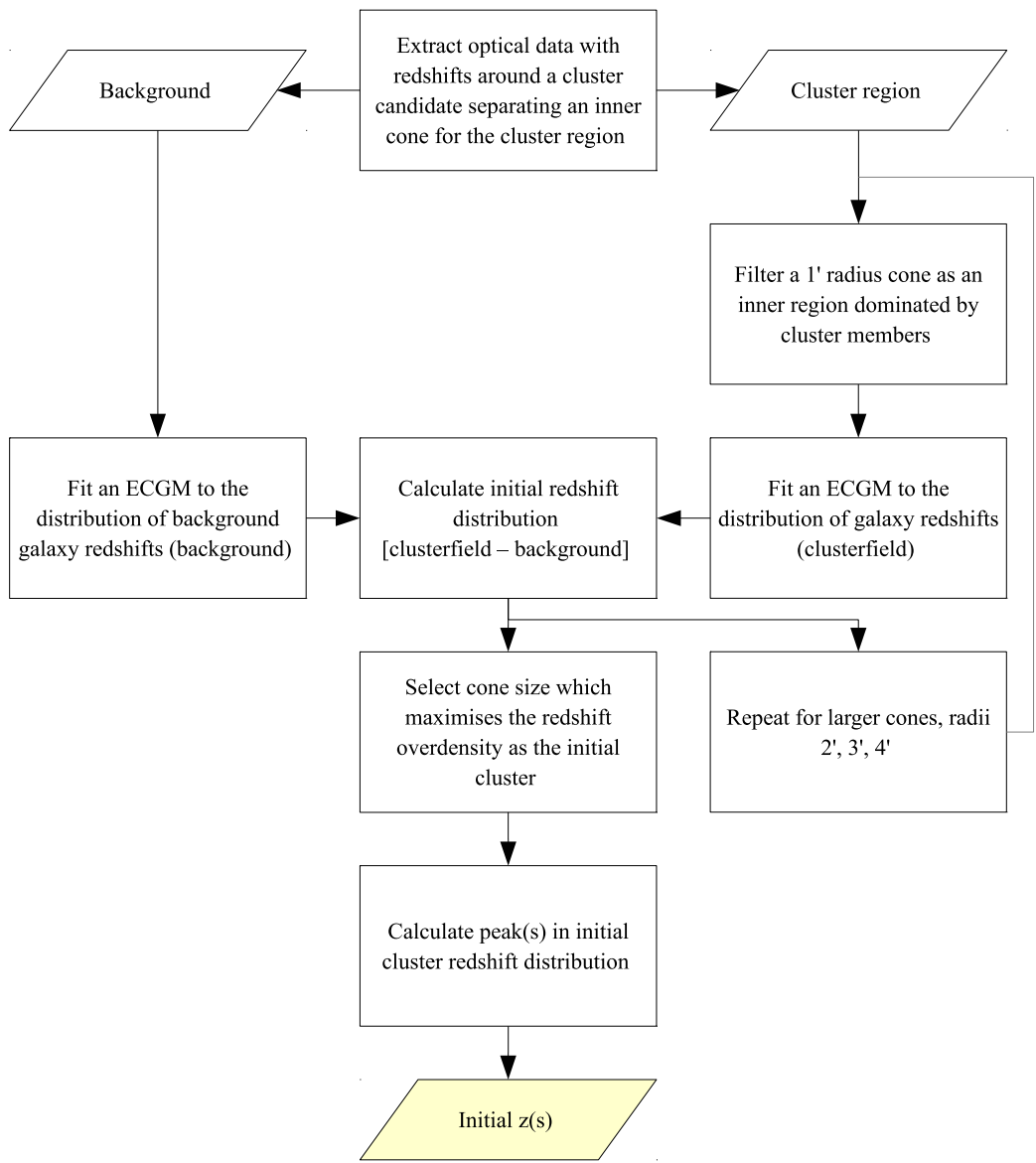} 
  \caption[A detailed flowchart showing the procedures of GMPhoRCC to determine an initial cluster redshift estimate.]{A detailed flowchart showing the procedures of GMPhoRCC to determine an initial cluster redshift estimate highlighted as procedure 1 in Figure \ref{fig:overviewflow}. }
  \label{fig:initialzflow}
\end{figure*}

Starting with redshift, Figure \ref{fig:initialzflow} expands procedure 1 of Figure \ref{fig:overviewflow} showing in detail the procedure used to arrive at an initial estimate for an inner cluster region. The redshift distribution of the inner cluster is modelled by taking the mixture model from a series of cones across a range of radii, $1$ - $4$ arc minutes and subtracting the background model. The inner cluster radius is then selected to produce the largest sum of the amplitudes of the peaks in the distribution. In addition to ensuring peaks can still be found in the case of miss-centring, this gives preference to regions producing multi-modal distributions where each peak can be subsequently analysed and used to assist with the characterisation. While this could be found with an investigation of the radial profile, this method is less sensitive to issues with overlapping clusters which indeed can be common ($\sim 20 $ percent of the GMBCG catalogue has a neighbour within $3$ arc minutes).

For a rigorous treatment of multi-modal redshift distributions, a secondary peak is investigated as a potential cluster provided the amplitude is at least $20$ percent of the primary. This threshold allows the analysis of potential structure without exploring noise or low level fluctuations in the distribution. Considering multiple peaks in this way occurs throughout GMPhoRCC resulting in a potentially large number of possible candidates from which the cluster is selected.

\subsubsection{Initial Colour Estimate}

\begin{figure*}
  \centering
  \includegraphics[width=373pt]{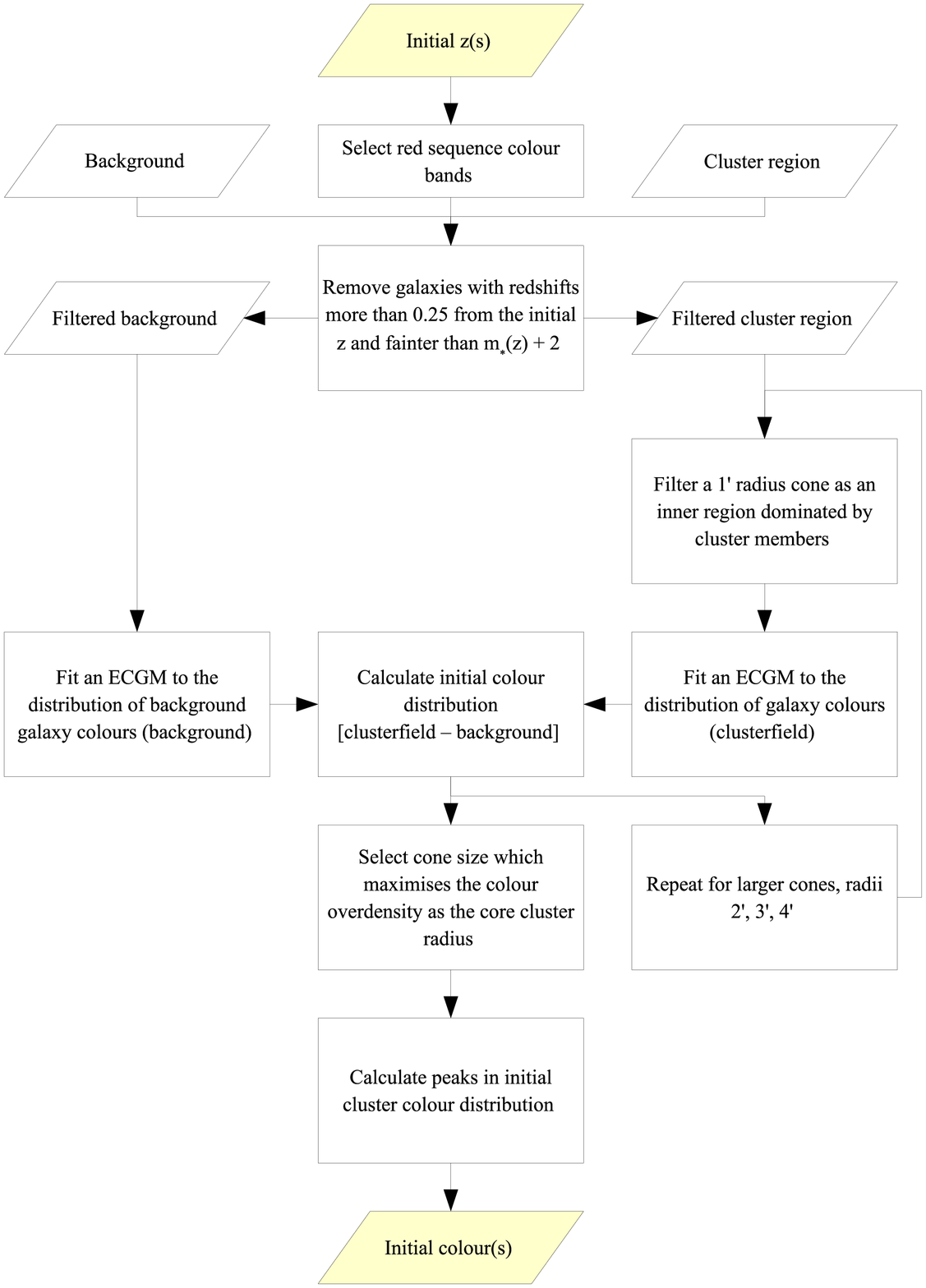} 
  \caption[A detailed flowchart showing the procedures of GMPhoRCC to determine an initial red sequence colour estimate.]{A detailed flowchart showing the procedures of GMPhoRCC to determine an initial red sequence colour estimate, highlighted as procedure 2 in Figure \ref{fig:overviewflow}.}
  \label{fig:initialcolourflow}
\end{figure*}

With the initial redshift estimate, an initial colour estimate for the inner cluster region proceeds as shown in Figure \ref{fig:initialcolourflow}. First an appropriate colour band is selected based on the initial redshift in line with \cite{gmbcg} shown in Table \ref{table:zcolourrange}. These values ensure that the main spectral feature of red sequence galaxies, the $4000$\r{A} break, remains in the band at a given redshift. This is important as this ensures the strongest colour clustering and contrast against the background. Additionally the redshift values overlap to account for possible failures of the initial estimate and help clusters where the $4000$\r{A} break sits between bands. 

\begin{table}
  \centering
  \caption[The most suitable red sequence colour bands for the initial redshift estimate.]{The most suitable red sequence colour bands for the initial redshift estimate. These values overlap to account for uncertainty in the initial redshift and those close to a transition.}
  \label{table:zcolourrange}
  \resizebox{!}{!}{
  \begin{tabular}{c l}
    \hline\hline
    Red sequence band & redshift range \\
    \hline
    $g-r$ & $0.0  \leq  z < 0.5$ \\
    $r-i$ & $0.3 \leq  z < 0.8$ \\
    $i-z$ & $0.6 \leq  z $ \\
    \hline
  \end{tabular}}
\end{table}

Before estimating an initial colour, galaxies are removed from the background and the cluster region which do not conform with the initial redshift; specifically those where the galaxy redshift is more than $0.25$ from the initial estimate and those fainter than would be considered in such a cluster. The faint end cut is taken as $m_*(z) + 2$ where the redshift dependent $m_*$ is taken from \cite{gmbcg} which was derived from the luminosity function of field galaxies. Initial red sequence colour estimation proceeds in a similar manner as redshift, where the colour estimate is taken as the peak in the background-subtracted colour distribution of an inner cluster region. The inner cluster region is determined again by considering cones with a range of radii and selecting the cone which maximised the sum of the peak amplitudes. This region is used as the inner cluster region in subsequent analysis.

\subsubsection{Red Sequence CMR}

\begin{figure*}
  \centering
  \includegraphics[width=246pt]{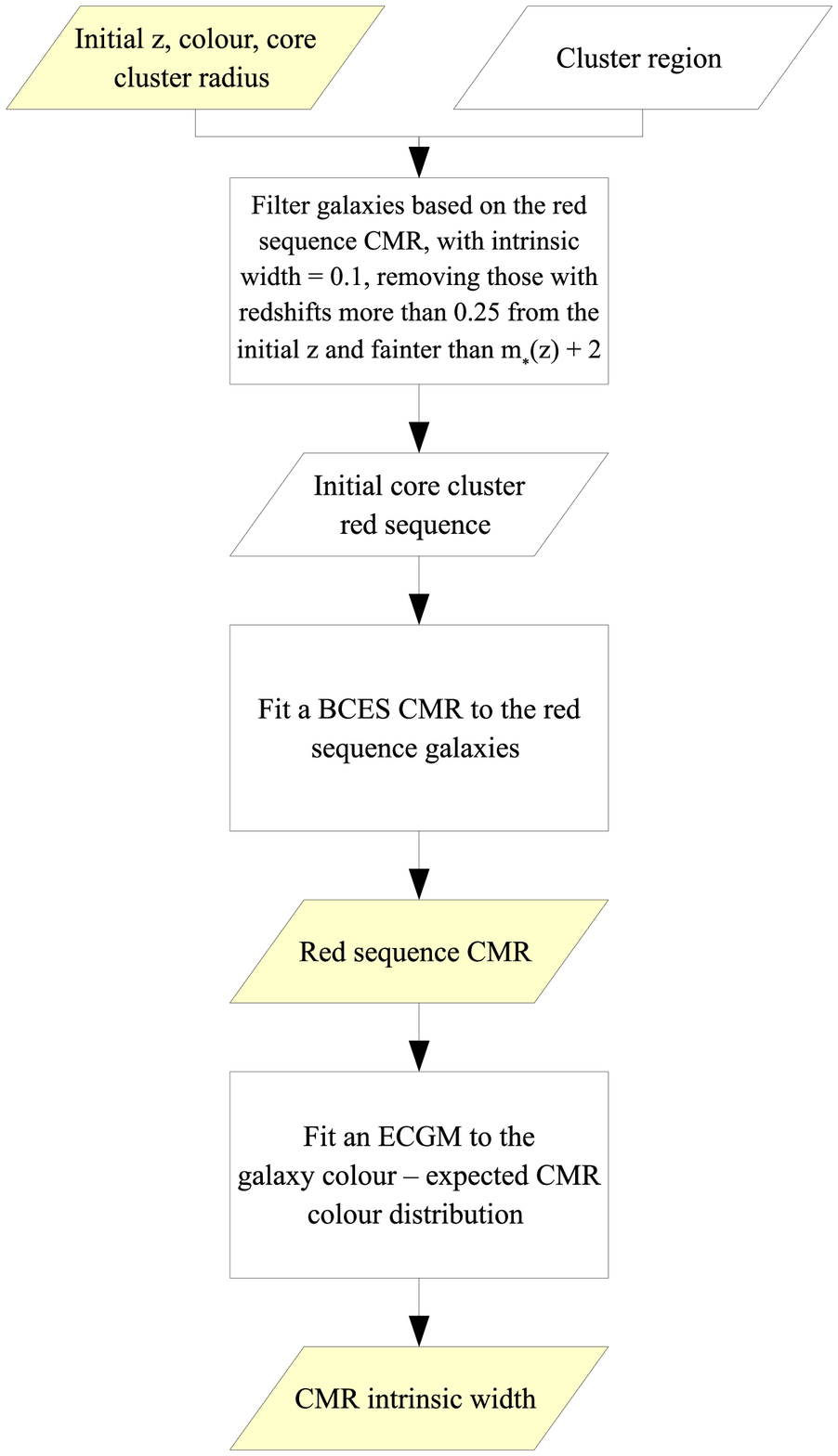} 
  \caption[A detailed flowchart showing the procedures of GMPhoRCC to filter remaining contamination and model the red sequence CMR.]{A detailed flowchart showing the procedures of GMPhoRCC to filter remaining contamination and model the red sequence CMR, highlighted as procedure 3 in Figure \ref{fig:overviewflow}.}
  \label{fig:cmrflow}
\end{figure*}

The red sequence CMR is determined using the initial estimates of redshift, colour and inner cluster radius as shown in Figure \ref{fig:cmrflow}. First galaxies from the inner region are filtered in a manner similar to that used to identify red sequence galaxies; all galaxies within $2\sigma$ of the initial colour estimate are kept for further analysis, where 
\begin{equation}
\sigma^2 = \sigma_{colour}^2 + \sigma_{RS}^2
\end{equation}

\noindent A broad initial red sequence width is used with $\sigma_{RS}=0.1$ to ensure only field contamination is removed.

With the red sequence dominating the remaining galaxies from the inner region, fitting a CMR proceeds using the bivariate correlated errors and intrinsic scatter (BCES) method \citep{bces}. This extends the standard least-squares method to account for intrinsic scatter and potentially correlated errors in both the dependent and independent variables. Additionally, the intrinsic width of the red sequence is determined from the distribution of colours around the CMR. 

In practice the distribution of galaxies around the CMR found using BCES takes the form of a Gaussian. By correcting for the slope and using a single component, the width of the error-corrected Gaussian mixture gives a good estimate of the intrinsic scatter of the red sequence.

\subsubsection{Red Sequence Redshift} 

Having isolated the red sequence from the background and inner cluster region using the determined CMR with measured intrinsic width with the $2\sigma$ approach, the peaks in the background subtracted photometric redshift distribution provide the potential cluster redshift. Additional estimates such as spectroscopic or colour model redshifts can be added at this point if desired by the user. If spectroscopic redshifts are available, averaging these for the red sequence galaxies can provide a more reliable redshift estimator. Finally the BCG is identified as the brightest galaxy on the red sequence in the inner region. 

\subsubsection{Candidate Selection} 

With the possibility of multiple candidates and the frequent ambiguity of cluster selection, the final step, shown in Figure \ref{fig:zrsflow}, filters the results to produce a primary as the most likely cluster candidate and a secondary as the next likely possibility. To help this process the potential clusters are first filtered to ensure the initial redshift, red sequence redshift and BCG redshift are all appropriate for the colour band used. Reducing the redshift overlap in Table \ref{table:zcolourrange} helps to remove the same candidates analysed in multiple bands. If this removes all the potential clusters the filter is not applied and the selection process continues.

\begin{table}
  \centering
  \caption[A tighter redshift-band relation for the red sequence than presented in Table \ref{table:zcolourrange}.]{A tighter redshift-band relation for the red sequence than presented in Table \ref{table:zcolourrange}. The reduced overlap helps to remove the same candidates analysed in multiple bands.}
  \label{table:zcolourrange2}
  \resizebox{!}{!}{
  \begin{tabular}{c l}
    \hline\hline
    Red sequence band & redshift range \\
    \hline
    $g-r$ & $0.00  \leq  z < 0.45$ \\
    $r-i$ & $0.35 \leq  z < 0.75$ \\
    $i-z$ & $0.65 \leq  z $ \\
    \hline
  \end{tabular}}
  
\end{table}

The remaining candidates are then ranked, based first on the consistency of the three main redshift estimators, initial, red sequence and BCG. Four cleanness bands are introduced shown in Table \ref{table:cleanli} where the most desirable candidates have the highest value.

\begin{table}
  \centering
  \caption[A list of the cleanness bands used to rank potential clusters.]{A list of the cleanness bands where the most desirable candidates have the highest band value. Redshifts are considered consistent with the colour band with they agree with Table \ref{table:zcolourrange2}.}
  \label{table:cleanli}  
  \begin{tabular*}{240pt}{c p{180pt} }
    \hline\hline
    cleanness & description  \\
    band  \\
    \hline
    $1$ & One or more of the main redshift estimators has not been found. \\
    $2$ & The red sequence and BCG redshift disagree by more than $0.1$ \\
    $3$ & All three redshift estimators are not consistent with the colour band\\
    $4$ & All remaining candidates. \\
    \hline
  \end{tabular*} 
\end{table}

\begin{figure*}
  \centering
  \includegraphics[width=378pt]{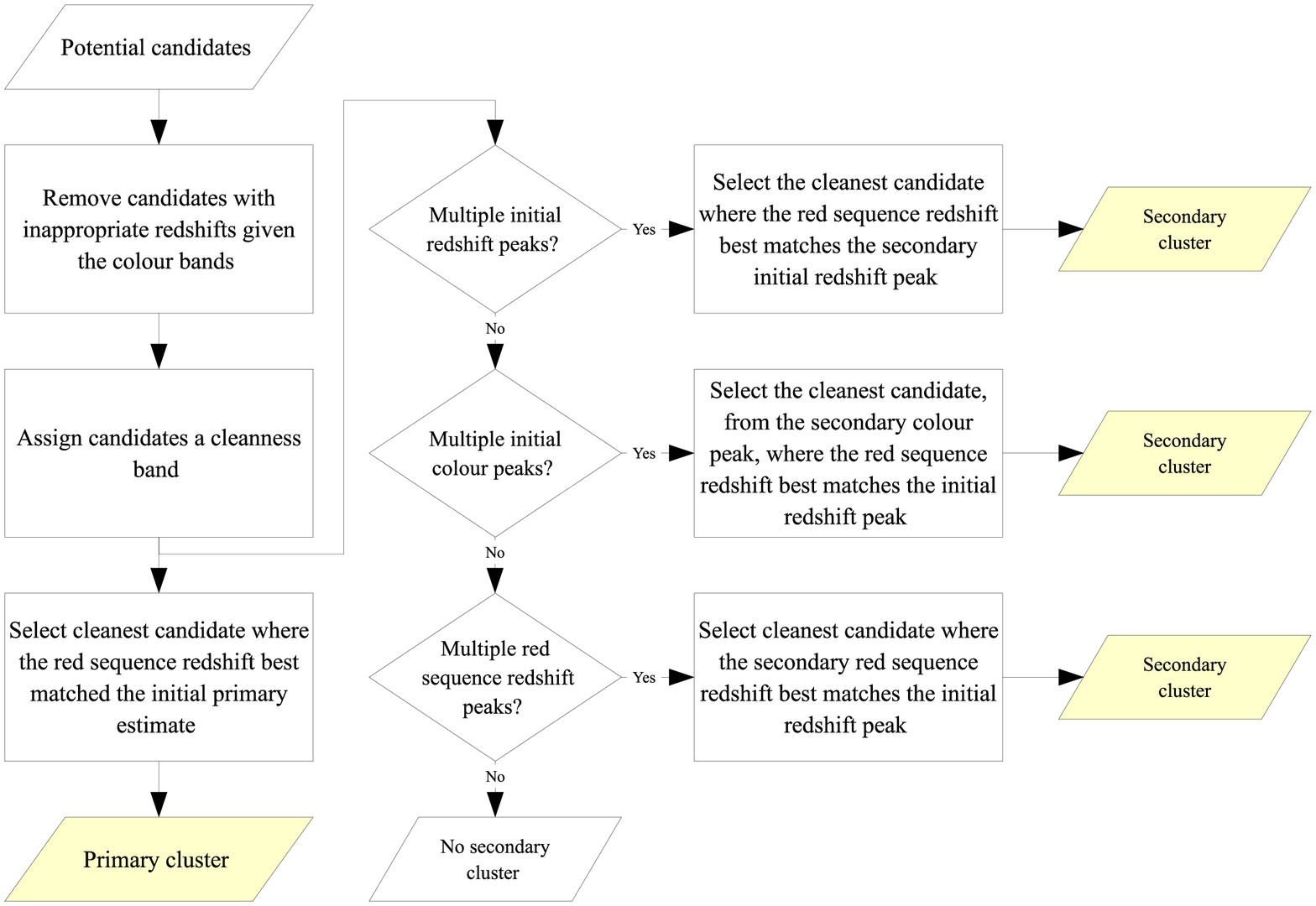} 
  \caption[A detailed flowchart showing the procedures of GMPhoRCC to select a primary and secondary cluster from a list of potential candidates.]{A detailed flowchart showing the procedures of GMPhoRCC to select a primary and secondary cluster from a list of potential candidates.}
  \label{fig:selflow}
\end{figure*}

\begin{figure*}
  \centering
  \includegraphics[width=382pt]{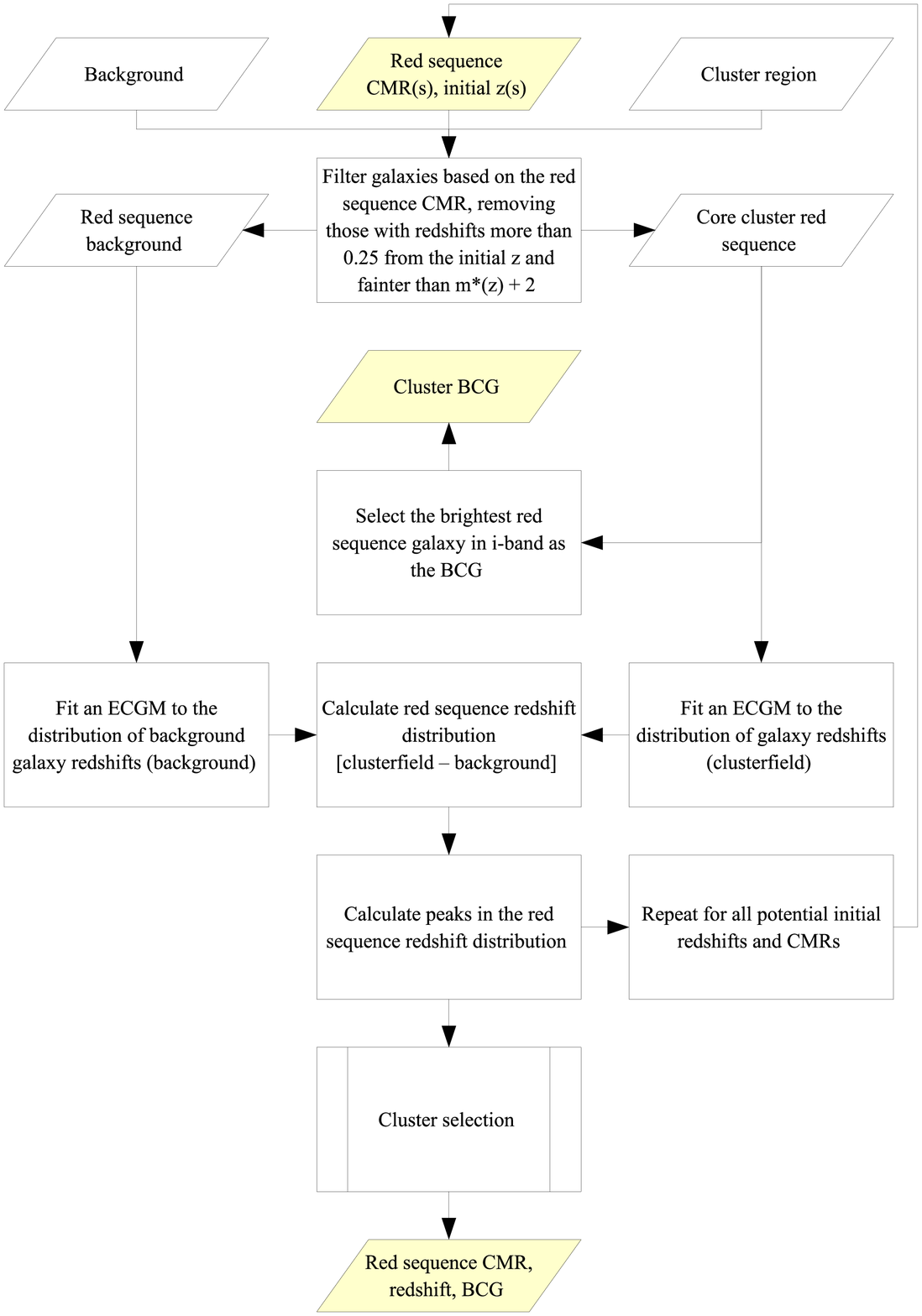} 
  \caption[A detailed flowchart showing the procedures of GMPhoRCC to estimate the cluster redshift from the red sequence CMR.]{A detailed flowchart showing the procedures of GMPhoRCC to estimate the cluster redshift from the red sequence CMR, highlighted as procedure 4 in Figure \ref{fig:overviewflow}. The cluster selection process selects a primary and secondary cluster as outlined in Figure \ref{fig:selflow}.}
  \label{fig:zrsflow}
\end{figure*}

These consistency checks help to remove candidates which may not represent clusters but rather, random enhancements in the background or foreground. Finally, to further rank clusters and break degeneracy, those where the red sequence redshift best matches the initial estimate provides the best selection, reducing the chance that the best candidates are spurious. The primary clusters are simply the cleanest candidates which best match the primary initial redshift estimate. A secondary cluster is assigned as the cleanest candidate associated with the earliest secondary peak (i.e. initial secondary redshifts considered first etc.) which best matched the initial estimate. This selection procedure is shown in detail in Figure \ref{fig:selflow}. 

While comparing the GMPhoRCC estimates to previously characterised spectroscopic clusters it is found that on average $35$ percent of targets characterised have an associated secondary cluster and of these only $13$ percent ($5$ percent of the total) better matches spectra than the primary. Again it has not only been shown that dealing with multi-modal distributions is necessary but also that the selection process of GMPhoRCC is able to reliably select the most appropriate characterisation from the potential candidates for the cluster.

\subsection{Richness}

GMPhoRCC measures richness as the number of red sequence galaxies, defined by the $2\sigma$ filtering, within a given radius fainter than the BCG and brighter than some redshift dependent cut-off, $m_*(z) + 1$. This takes the form of the maxBCG \citep{maxbcg} and GMBCG \citep{gmbcg} richness which ensures the red sequence $i$-band magnitude range is consistent as a function of redshift. In agreement with maxBCG, GMBCG and redMaPPer \citep{redmapper} $m_*$ is taken from the luminosity function of field galaxies determined by \cite{blanton}.

For consistency across a range of cluster sizes, GMPhoRCC considers $n_{200}$, the richness inside the characteristic radius $r_{200}$. As measuring $r_{200}$ directly is only possible with gravitational lensing, an intermediate $0.5$h$^{-1}$Mpc fixed aperture richness, $n_{gals}$, is used following the analysis of maxBCG and GMBCG, from which $r_{200}$ is found. Using the maxBCG clusters and the weak lensing derived $r_{200}$ - $n_{200}^{maxBCG}$ scaling relation from \cite{hansen2009} the following scaling relation was found by binning clusters by $n_{gals}$ and fitting $r_{200}$. Direct derivation of this relation using weak lensing $r_{200}$ will reduce the scatter in this relation and is left for future work.
\begin{equation}
\label{eq:scale}
r_{200} = 0.237 (n_{gals})^{0.4}
\end{equation}

In addition to counting galaxies, richness is also estimated using the luminosity function method of \cite{high}. This involves fitting a luminosity function within a magnitude range where the photometry is believed to be complete then integrating up to $m_* + 1$. Finding this range is simply done by inspecting a magnitude histogram where a limit can be assigned after which the density drops with increasing magnitude. Rather than using the binning approach to fit a Schechter function a new probabilistic approach has been developed which gives more reliable fits across a range of cluster richnesses. With appropriate normalisation the luminosity function, Equation \ref{eq:shec2}, gives the number of galaxies within the magnitude range $m \rightarrow m+dm$ and hence the probability that a galaxy has a particular magnitude given the parameters of the Schechter function can be approximated by Equation \ref{eq:lumprob}.
\begin{eqnarray}
\label{eq:shec2}
\phi(m,\theta) \mathrm{d} m&=&0.4\lbrack\ln(10)\rbrack\phi_*10^{-0.4(m-m_*)(\alpha +1)}\nonumber \\
&\times& \exp[-10^{-0.4(m-m_*)}] \mathrm{d} m
\end{eqnarray}

\begin{equation}
\label{eq:lumprob}
P(m | \theta)=\frac{\phi(m,\theta) \mathrm{d} m}{\textrm{Total Number of Galaxies}}
\end{equation}

\noindent where $\theta$ represents the parameters of the Schechter function, namely $\phi_*$, $m_*$ and $\alpha$. Using Bayes' theorem, the likelihood of the parameters, $\mathcal{L}$, is given by combining the probabilities from all galaxies.
\begin{equation}
\label{eq:lumlike}
\mathcal{L} = \prod_{k=1}{P(\theta |m_k)}  = \prod_{k=1}{\frac{P(m_k | \theta)P(\theta)}{P(m_k)}}
\end{equation}

\noindent The cluster luminosity function is then defined by the parameters which maximise this likelihood. By assuming flat priors this is equivalent to minimising the log-likelihood shown in Equation \ref{eq:lumll}. 
\begin{equation}
\label{eq:lumll}
\ln(\mathcal{L}) \propto \sum_{k=1}{\ln{\phi(m_k,\theta)}}
\end{equation}

\noindent In addition to using the \cite{high} fixed faint end ($\alpha=-1$), constraining the parameters to satisfy Equation \ref{eq:lumcon} greatly increases the reliability of the fit:
\begin{equation}
\label{eq:lumcon}
\int{\phi(m,\theta) \mathrm{d} m} = \textrm{Total Number of Galaxies}
\end{equation}

\noindent This ensures a reasonable luminosity function is recovered which, when integrated across the previously determined magnitude range, returns the total number of galaxies observed.

Minimising Equation \ref{eq:lumll} subject to the constraint shown in Equation \ref{eq:lumcon} proceeds using the standard sequential least squares method described by \cite{slsqp}.

Although an improvement this method still produces unreliable results for very low numbers of galaxies hence, when fitting $5$ or fewer data points, $m_*$ is fixed based on the cluster redshift and the luminosity function of field galaxies determined by \cite{blanton}.

\begin{figure*}
  \centering
  \includegraphics[width=362pt]{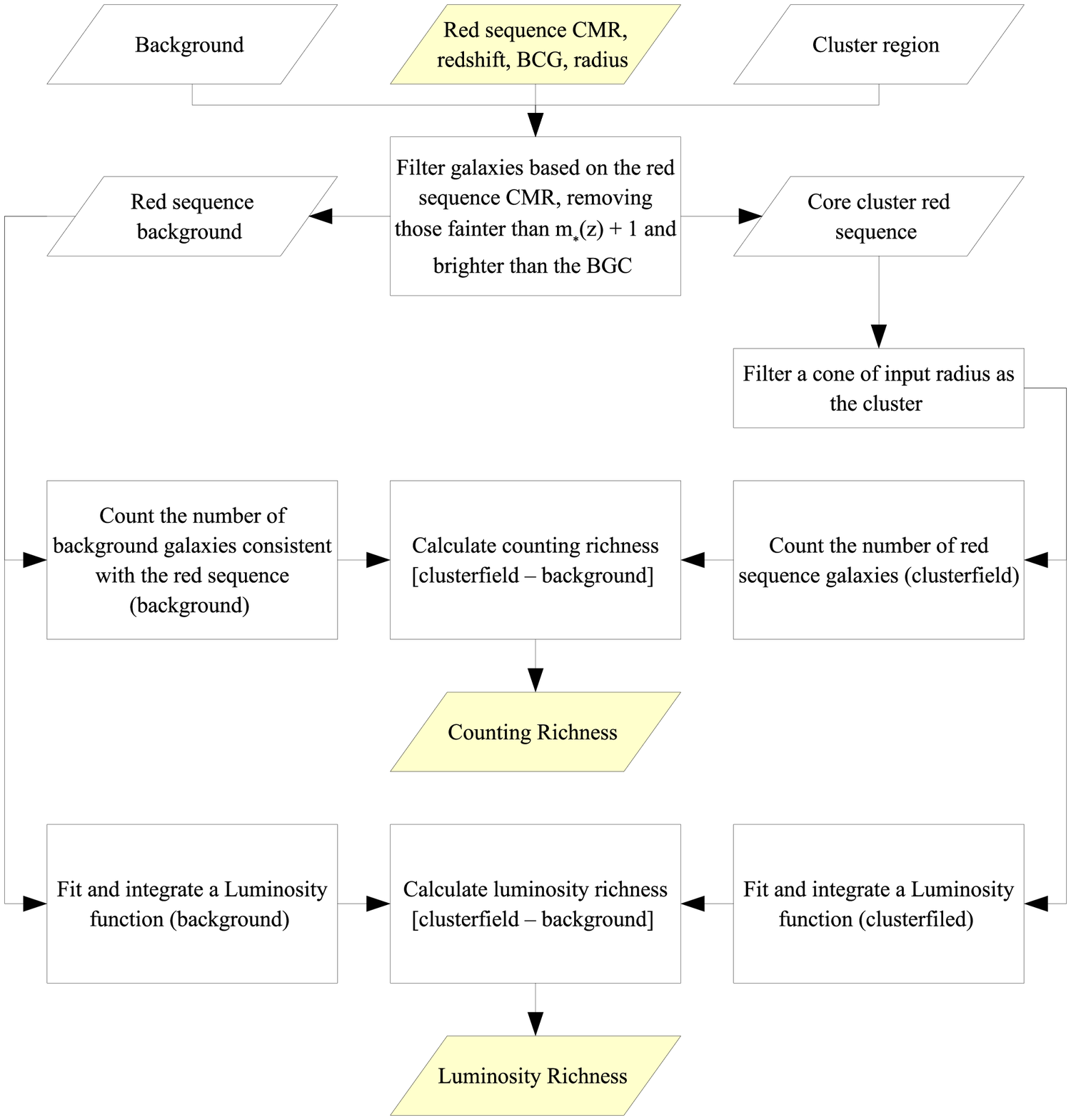} 
  \caption[A detailed flowchart showing the procedures of GMPhoRCC to estimate the cluster richness.]{A detailed flowchart showing the procedures of GMPhoRCC to estimate the cluster richness, highlighted as procedure 5 in Figure \ref{fig:overviewflow}. An input radius of $0.5$h$^{-1}$Mpc  leads to the intermediate richness, $n_{gals}$ with $n_{200}$ given by $r_{200}$.
  }
  \label{fig:richflow}
\end{figure*}

Combining these methods, Figure \ref{fig:richflow} expands procedure 5 from Figure \ref{fig:overviewflow} in more detail showing the steps taken to estimate cluster richness. By using an input radius of $0.5$h$^{-1}$Mpc the intermediate richness, $n_{gals}$ is determined, with $n_{200}$ found by using $r_{200}$.

\subsection{Optical Data}

Although designed for use with any optical data, initial calibration and development of GMPhoRCC has been driven with the use of optical data from the Sloan Digital Sky Survey. The tenth data release presented in \cite{sloan10}, provides coverage of $14,555$ squared degrees in the northern hemisphere giving $95$ percent completeness down to $21.3$ magnitudes in $i$-band giving $\sim 90$ million suitable galaxies, $\sim 1.9$ million with spectra.

The input optical data were selected from the Galaxy view of PhotoObjAll table using the following query to ensure cleanness and completeness in photometry.

\begin{verbatim}
SELECT * from GALAXY 
WHERE

	(dered_i) < 21.0 AND

	(modelMagErr_g / dered_g) < 0.1 AND
	(modelMagErr_r / dered_r) < 0.1 AND
	(modelMagErr_i / dered_i) < 0.1 AND
	
	(insidemask)=0 AND
	(clean)=1 AND
	(extinction_i) < 0.5
	
	(dered_g - dered_r) BETWEEN -1.5 and 5 AND
	(dered_r - dered_i) BETWEEN -1.5 and 4 AND
	(dered_i - dered_z) BETWEEN -1.5 and 4 AND
\end{verbatim}

\noindent The $10$ percent cut on colour errors with extinction and masking constraints, as used by \cite{gmbcg}, ensures the optical data are clean which greatly improves the Gaussian mixture fitting procedure. In addition the colour cuts and $i$-band constraint helps to remove extreme objects with likely erroneous photometry which adversely bias the Gaussian Mixture models of the cluster candidates. While the $i$-band cut is specific to the SDSS, the colour and error constraints are recommended for GMPhoRCC regardless of the source of the optical data to ensure clean photometry.

In addition to multi-band photometry GMPhoRCC makes use of photometric redshifts rather than using assumed colour-redshift relations. Within the SDSS DR10, the PhotozRF table provides the most suitable redshifts, calculated using the random forest regression technique of \cite{randomforests}. While not essential, these provide well understood Gaussian errors which are ideal for use with the error-corrected Gaussian Mixture models of Section \ref{sec:ecem}.

\subsection{Quality Control}

One of the main goals of GMPhoRCC is to provide a means of quality control to help identify possible catastrophic failures. As part of this many flags have been introduced to trace how clusters propagate through the algorithm. These flags trace potential issues with fits, multi-modal distributions and inconsistent redshifts with a full list given in Appendix \ref{sec:gmphorccoutputs}.

Although a large source of ambiguity and potential failure results from the presence of multi-modal distributions, with the prevalence of these (seen in $\sim 70$ percent of clusters) and the success of the candidate selection shown in Figure \ref{fig:selflow}, these are not sufficient to identify catastrophic failures alone. The strongest indicators of failure however, are the presence of inconsistent redshifts or low richnesses.

Considering low richness, firstly this indicates that the distribution modelling may be unreliable, fitting many parameters to only a few data points. More importantly this could indicate an issue with the red sequence, either the candidate cannot be optically confirmed as a cluster or the red sequence has been missed altogether. In either case this is the strongest indication of catastrophic failure.

Large discrepancies between the red sequence and the BCG is also a strong indicator of catastrophic failure particularly with regards to redshift. Large discrepancies in redshift (larger than expected considering measurement error), can indicate either a problem with red sequence modelling, BCG selection or cluster redshift. 

By combining these flags, quality markers are assigned to clusters as an indicator of the reliability of the optical characterisation, shown in Table \ref{table:markers}. While these quality markers, $q$, are universal it is noted that the redshift inconsistency flags, \verb|INCONSISTENT_Z_PHOT| and \verb|INCONSISTENT_Z_SPEC|, assigned for cases where $|z_{RS} - z_{BCG-phot}| > \Delta z_{cp}$ and  $|z_{RS} - z_{BCG-spec}| > \Delta z_{cs}$ (see Table \ref{table:markers1}), must be calibrated for specific sources of the photometric redshifts. Here $\Delta z_{cp}$ and $\Delta z_{cs}$ are respectively photometric and spectroscopic redshift consistency bounds where, for the SDSS DR10, $\Delta z_{cp} = 0.035$ and $\Delta z_{cs}=0.025$. With less reliable photometric redshifts these should be relaxed to larger bounds.

\begin{table}
  \centering
  \caption[A list of the quality markers assigned to clusters based on the GMPhoRCC flags.]{A list of the quality markers, $q$, assigned to clusters based on the GMPhoRCC flags.}
  \label{table:markers}
  \begin{tabular*}{240pt}{c l }
    \hline\hline
    $q$ &     Description \\
   \hline
    $-1$ & no optical coverage \\
    $0$ & no characterisation found \\
    $1$ & $n_{200}<1$, large redshift inconsistencies, masking issues  \\
    $2$ & $n_{200}<3$, small redshift inconsistencies   \\
    $3$ & clean \\
    \hline
  \end{tabular*}
\end{table}

With these quality markers the characterisations can be separated into various quality subsets, as shown in Table \ref{table:qualsub}; `clean' with $q\geq3$ representing the cleanest set with most problem clusters removed; `mid' with $q\geq2$, a middle subset with only the worst clusters removed; and `detection' with $q\geq1$, the full list of clusters considered to have been detected.  

\begin{table}
  \centering
  \caption[A list of the subsets of clusters based on the GMPhoRCC quality marker.]{A list cluster subsets based on the GMPhoRCC quality marker, $q$, used remove potentially erroneous characterisations.}
  \label{table:qualsub2}
  \begin{tabular*}{240pt}{l c p{155pt}}
    \hline \hline
    Subset      &   $q$   & Description\\
    \hline
    Detection   & $\geq 1$ &  All clusters considered to have been detected \newline i.e. estimates were found for both redshift and richness \\
    
    Mid         & $\geq 2$ &  An intermediate subset removing the worst outliers \newline i.e. removing clusters with very low richness or large discrepancies between redshift estimates \\
    
    Clean       & $\geq 3$ &  The cleanest subset removing the majority of outliers \newline i.e. removing cluster with low richness and discrepancies between redshift estimates \\
    \hline
  \end{tabular*}
\end{table}

\subsection{Computational Performance}

GMPhoRCC is aimed primarily for use with standard desktop computers and as such does not require substantial computational resources. Development has proceeded using \verb|python| $2.7.3$ with the \verb|scipy|\footnote{\url{http://www.scipy.org/}} module providing many of the mathematics routines, particularly the sequential least squares method used to fit luminosity functions. GMPhoRCC experiences two main bottlenecks, first from the retrieval of the optical data either from a database or local files and secondly from fitting Gaussian mixtures. While little can be done with the data retrieval, the Gaussian mixture fitting is developed using \verb|Fortran| $90$ which provides a factor of $10$ speed improvement over native \verb|python| and is twice as fast as the \verb|c++| version employed by \cite{gmbcg}. The final performance improvement comes from the utilization of multiple threads available in even the most basic computers. While GMPhoRCC does not implement full parallelisation at the \verb|Fortran| level, the \verb|Parallel| \verb|Python|\footnote{\url{http://www.parallelpython.com/}} module allows for several cluster candidates to be analysed simultaneously. Although more were available little improvement was found beyond six threads due to restrictions in the retrieval of the optical data.

As an example of typical performance, $6$ threads from an Intel $3770$k $4.2$GHz processor with $16$GB of PC3-$19200$ RAM, accessing the optical data locally from a hard disk has a characterisation time of $42$ seconds per cluster per thread allowing the full characterisation of the XCS catalogue, $503$ clusters, within $59$ minutes.

\section{Evaluation}
\label{sec:mocks}

Evaluation of GMPhoRCC proceeds with a two-prong approach, first by comparing characterisations with other algorithms using spectroscopic clusters and secondly by investigating mock galaxy clusters. In addition to driving the development process, particularly the calibration of the quality control system, these comparisons allow for detailed understanding of the GMPhoRCC optical selection function.

\subsection{Comparison with existing catalogues}

Comparisons with existing catalogues proceeded using spectroscopic clusters selected from the GMBCG \citep{gmbcg}, NORAS \citep{noras}, REFLEX \citep{reflex} and XCS \citep{xcsoptical} catalogues. As richness measures are specific to the exact form of the algorithm and optical data, evaluation of the GMPhoRCC richness is thus deferred to analysis with mock clusters where comparisons with `true' cluster values are possible.

\begin{figure*}
  \begin{minipage}{504pt}
    \includegraphics[width=240pt]{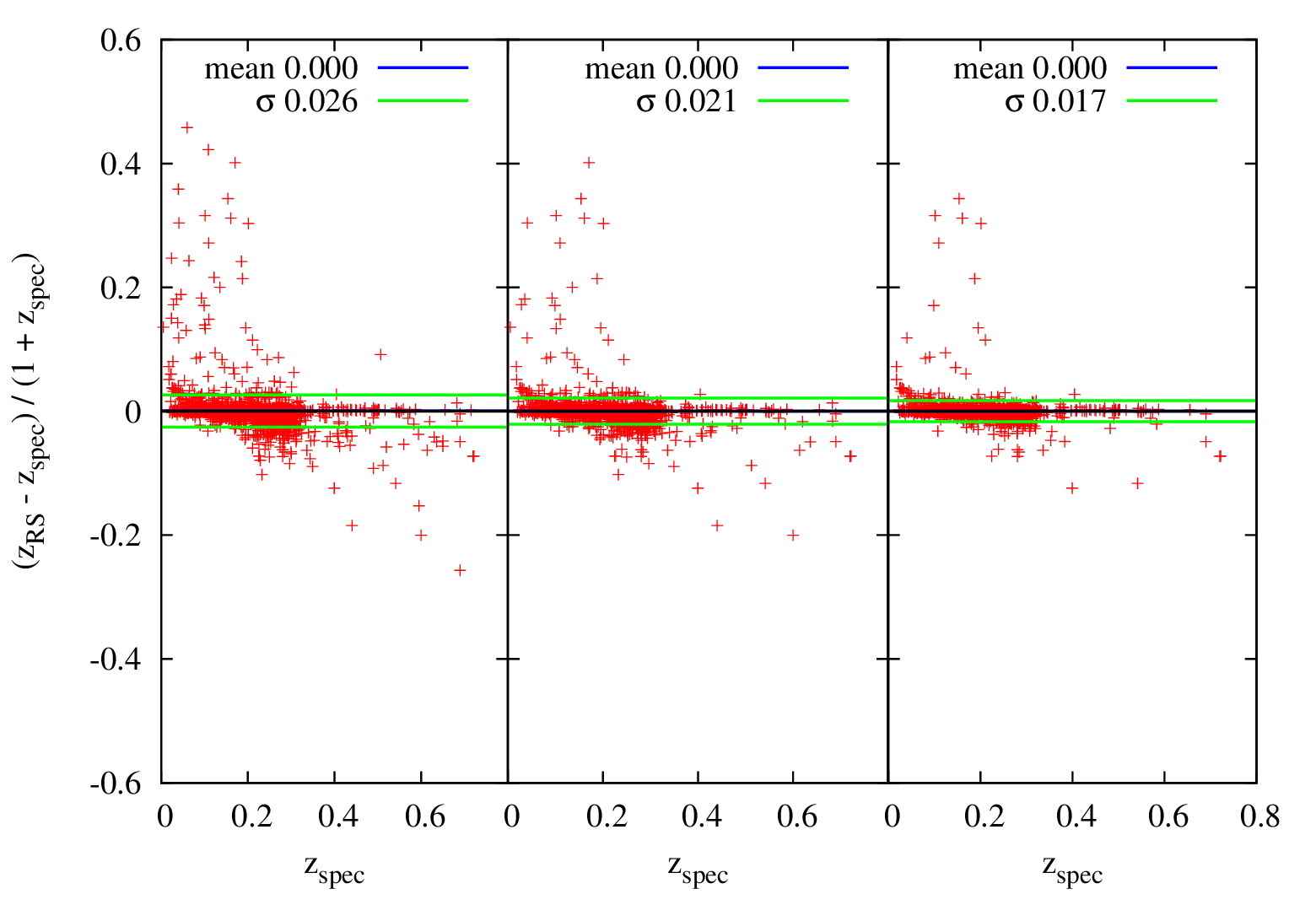}\hfill	
    \includegraphics[width=240pt]{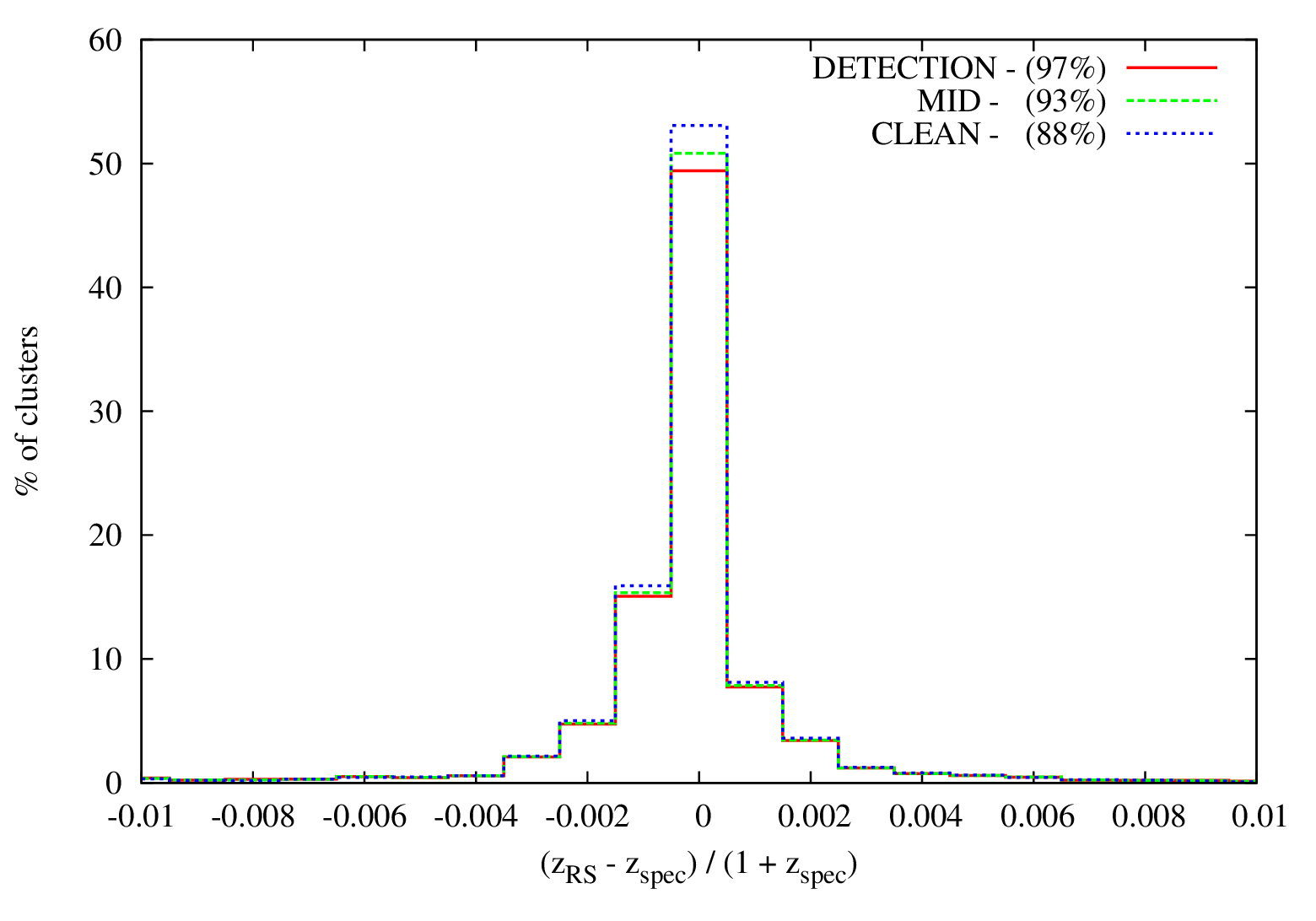} 
    \caption{A comparison of GMPhoRCC photometric red sequence redshifts to spectra using $4501$ clusters with DR10 coverage from GMBCG, NORAS, REFLEX and XCS. \textit{Left panel:} A Scatter plot highlighting the quality control, showing from left to right, the detection, mid and clean subsets. While some discrepancies remain the majority of outliers have been removed in the clean subset. Although they have been correctly identified as problems, very low redshift clusters ($z<0.05$) are not characterised well by GMPhoRCC due to poor contrast against the background and limitations in the field area. In addition high redshift clusters $z>0.5$ are subject to large discrepancies due to incompleteness and increasing photometric errors. 
		\textit{Right panel:} The distribution of redshift comparisons where the results has been normalised and split into the separate quality subsets where the legend shows the fraction of the total clusters in each set. While the majority of all estimates are within $|z_{RS} - z_{spec}| / (1 + z_{spec}) <0.01$, the clean subset can again be seen to have removed the worst estimates with a greater fraction attaining this bound.}
    \label{fig:cgmxscatter}
\end{minipage}
\end{figure*}

With a total of $706$ X-ray and $3795$ optically detected clusters with spectra, direct evaluation of the GMPhoRCC redshift estimate is possible. Of the $4501$ clusters, redshift estimates were found for $97.3$ percent and compared to spectra and shown in Figure \ref{fig:cgmxscatter}. Although some discrepancies are present the quality markers are shown to identify and remove the worst outliers. Additionally, while the majority of all estimates are within $|z_{RS}-z_{spec}| / (1 + z_{spec}) < 0.01$, the clean subset attains a larger fraction within this bound and less contamination with outliers. It is noted however that at low redshifts, $z<0.1$, many cluster estimates are erroneous where limitations in field area and poor contrast against the background result in cases where field galaxies dominate the cluster distributions making it difficult to isolate the red sequence. Incompleteness and increasing measurement errors in the photometry at high redshift again cause issues with the red sequence detection. In addition to these redshift limitations it expected that low richness clusters produce the most outliers, where it is more difficult to isolate and model the red sequence with a sparse number of galaxies.

\begin{figure*}
  \begin{minipage}{504pt}
    \includegraphics[width=240pt]{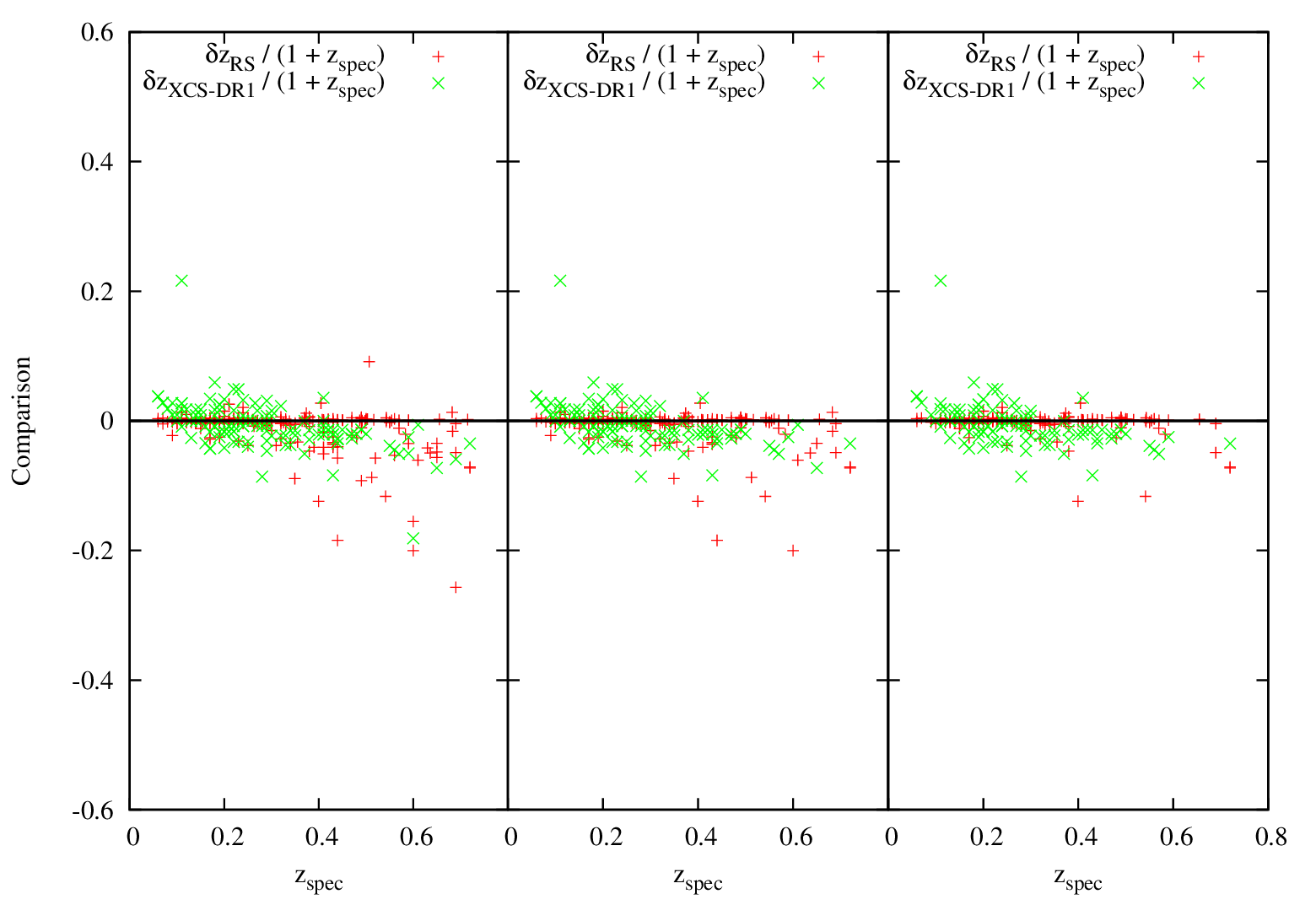}\hfill	
    \includegraphics[width=240pt]{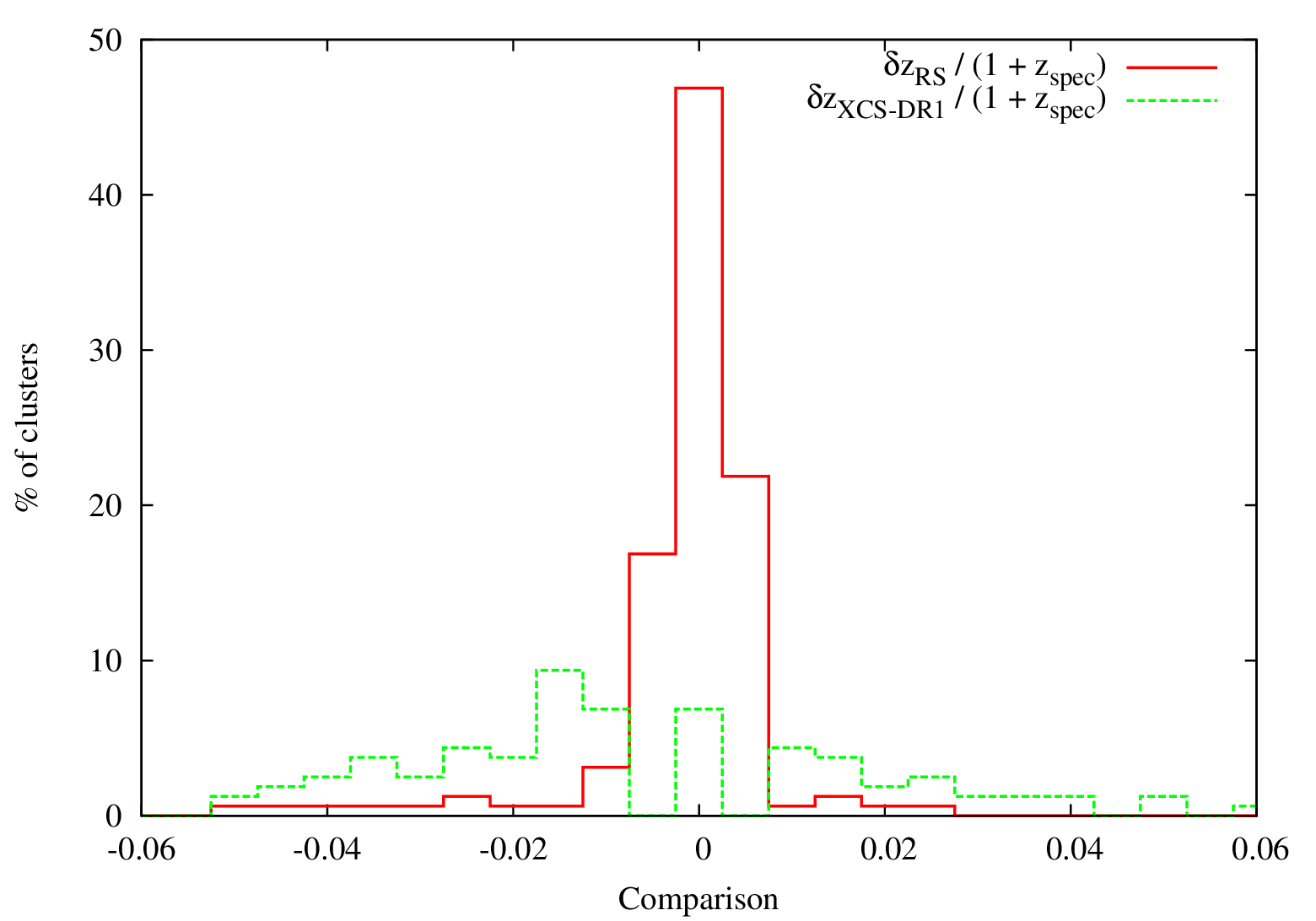} 
  \caption{Redshift comparisons for the GMPhoRCC and XCS photometric redshifts to spectra for a subset of $131$ XCS clusters where, $\delta z_{RS} = z_{rs} - z_{spec}$ and $\delta z_{XCS-DR1} = z_{XCS-DR1} - z_{spec}$. \textit{Left panel:} A scatter plot highlighting the quality control, showing from left to right the detection, mid and clean subsets. 
	\textit{Right panel:} The distribution of the redshift comparisons for the clean subset highlighting the substantial improvement offered by GMPhoRCC over XCS, providing more accurate estimates with a lower scatter around the spectroscopic redshift. Although the detection subset has a few more extreme outliers a much greater fraction than from XCS agree within $|z_{phot}-z_{spec}| / (1 + z_{spec}) < 0.01$, with the clean subset attaining the highest fraction in this band. In addition to providing more accurate redshifts, the estimates are independent of any colour-redshift model as employed by XCS. }
  \label{fig:xcs_z_rs_his}
\end{minipage}
\end{figure*}

In addition to comparisons with spectra, a subset of $131$ XCS clusters with both spectroscopic and photometric redshifts provide an excellent resource to compare the performance of GMPhoRCC and XCS. Figure \ref{fig:xcs_z_rs_his} shows the substantial improvement offered by GMPhoRCC, providing more accurate estimates with lower scatter around the spectroscopic redshifts. In addition to providing more accurate redshifts, the estimates are independent of any colour-redshift model as employed by XCS.

\subsection{Richness Scaling}

To assess the validity of the GMPhoRCC richness as a mass proxy, an initial investigation of richness scaling is explored for the x-ray clusters from the XCS catalogue. Of particular interest is the determination of X-ray - optical scaling relations, which, in work similar to \cite{xrayscale} and \cite{xrayscale2}, relies on the tight correlation between X-ray observables, such as temperature, to the cluster mass, in order to calibrate the GMPhoRCC richness as an optical mass proxy. While this paper illustrates the validity of the GMPhoRCC richness as a proxy for cluster properties, a complete analysis of such richness scaling is left for future work.

\begin{figure}
	\centering
	\includegraphics[width=240pt]{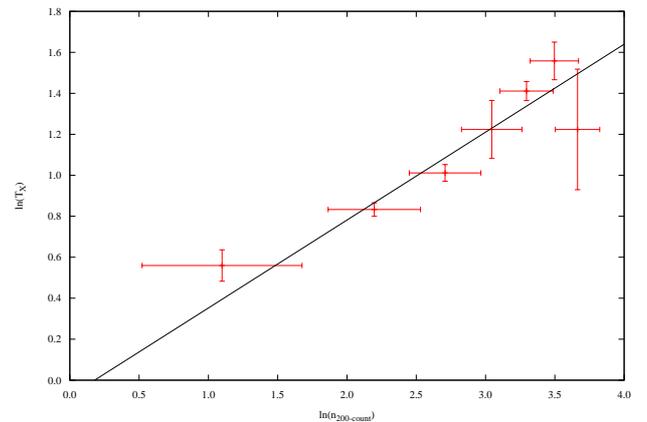}
	\caption[A preliminary analysis of the $T_{\mathrm{x}}$ - $n_{200}$ scaling relation produced using a clean spectroscopic subset of XCS.]{A preliminary analysis of the $T_{\mathrm{x}}$ - $n_{200}$ scaling relation produced using a clean spectroscopic subset of XCS. By stacking the clusters in richness bins and using the BCES method of Section \ref{sec:pipered} and \cite{bces}, a best fit linear model is found from the average temperatures and $n_{200}$. The best fit scaling relation of Equation \ref{eq:xscale} is shown with $\alpha = -0.08 \pm 0.08$, $\beta = 0.43 \pm 0.03$ and a scatter of $\sigma_{\ln T_\mathrm{x} | \ln n_{200}} = 0.14$ }.
		\label{fig:xrayscale}
\end{figure}

The previous subset of $131$ clean spectroscopic clusters from XCS are analysed to determine the correlation between GMPhoRCC richness and X-ray temperature, modelled as power law scaling relation similar to those used by \cite{xrayscale2} and defined below:

\begin{equation}
\label{eq:xscale}
\ln(T_{\mathrm{x}}) = \alpha +\beta \ln (n_{200})
\end{equation}

\noindent where $\alpha$ and $\beta$ are constants. Determination of this relation proceeds by stacking the clusters in richness bins and using the BCES method of Section \ref{sec:pipered} and \cite{bces}. While this is a rather simplistic approach for illustration, future analysis is indented using more sophisticated techniques, such as the Bayesian method of \cite{xrayscale2} and \cite{redmapper2}. Figure \ref{fig:xrayscale} demonstrates the $T_{\mathrm{x}}$ - $n_{200}$ scaling relation finding, $\alpha = -0.08 \pm 0.09$ and $\beta = 0.43 \pm 0.03$ with a scatter $\sigma_{\ln T_\mathrm{x} | \ln n_{200}} = 0.14$.

Again, as an illustration only, a clear correlation can be observed highlighting the validity of the GMPhoRCC richness as an optical proxy for cluster properties.

\subsection{SDSS-like Mocks}

While comparisons with existing catalogues are a useful tool to evaluate a characterisation method these can only take us so far. Existing methods are subject to their own strengths, weaknesses and selection functions hence comparisons with a controlled `truth' are considered with the use of mock galaxy clusters. Mock clusters can be constructed with known redshifts, richnesses and CMRs either through simulations (\citealp{orcamock}, \citealp{orca}, \citealp{highmock}, etc.) or empirically (\citealp{maxbcg}, \citealp{gmbcg}, etc.). 

SDSS-like empirical mocks are constructed for use with GMPhoRCC by adding artificial clusters to field galaxies, derived from existing cluster detections and and SDSS optical data. This has the advantage of producing mocks tailored to match the available photometry allowing the specific evaluation of GMPhoRCC for the optical data. 

Artificial clusters were generated by resampling galaxies from existing red sequences and BCGs to reproduce five main aspects of real clusters.

\begin{enumerate}
\item A suitable BCG
\item Radial profile 
\item Redshift distribution
\item Luminosity function
\item CMRs / Colour distributions
\end{enumerate}

As these are dependent on the properties of the cluster it is necessary to resample from red sequences which best match the target mock. Rather than using a small number of well observed seed clusters, red sequences were identified and stacked in redshift/richness space in order to provide a source of galaxies suitable for a range of mock properties.

Using GMPhoRCC, $10,000$ very clean red sequences were identified from the C4, GMBCG, REFLEX, NORAS and XCS catalogues with very good agreement between spectroscopic and GMPhoRCC redshifts, $|z_{RS} - z_{spec}| < 0.005 $. By separately stacking the BCG and red sequence galaxies of these clusters in redshift/richness bins, a larger source is produced to sample cluster properties than from considering these individually. Stacking many clusters in this way ensures that each bin is dominated by the red sequence where the bulk properties are representative of a cluster with the bin redshift and richness. 

While the available richness were fixed by the original clusters, extrapolation by adding a fixed $\Delta z$ to each galaxy allowed a larger redshift range to be sampled. Photometry was then adjusted with K$+$e corrections to account for evolution and observations at different redshifts. K-corrections were performed using \verb|KCORRECT| v4.2 from \cite{blanton2} with evolutionary corrections taken from \cite{maxbcg}. Colour evolution models from \cite{lrgevolution} were also considered but provided no significant deviation from the main results of this section.

\begin{figure}
  \centering
  \includegraphics[width=240pt]{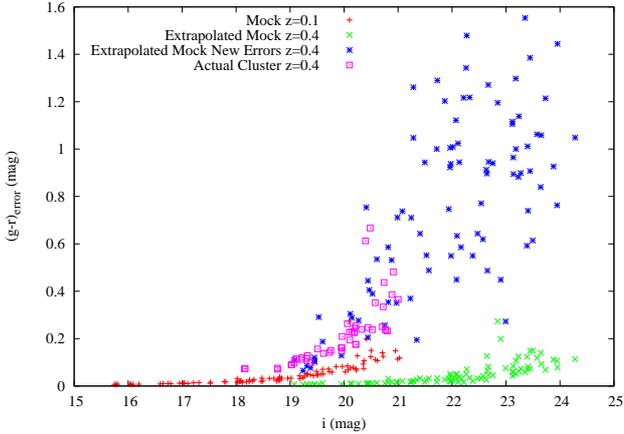} 
  \caption[A plot showing how the $g-r$ error compares between the $z=0.1$ seed cluster, the extrapolated cluster at $z=0.4$ with and without the new errors and a real $z=0.4$ cluster.]{A plot showing how the $g-r$ error compares between the $z=0.1$ seed cluster, the extrapolated cluster at $z=0.4$ with and without the new errors and a real $z=0.4$ cluster. Without the new photometric errors it is clear that the extrapolated cluster underestimates the $g-r$ error. The new errors are seen to be in good agreement with an observed cluster at the same high redshift. }
  \label{fig:mockerror}
\end{figure}

While extrapolating to much higher redshifts care is needed to reproduce appropriate errors. This mainly affects high redshift artificial clusters which should possess higher errors than the low redshift seed due to the fainter photometry. To reproduce appropriate errors, a sample of $\sim 500,000$ red sequence galaxies are used to model the various error distributions. For photometry errors the distribution is modelled as a function of band magnitude, with redshift considering the distribution as a function of $i$-band magnitude and redshift. For high redshift extrapolation, a new error is drawn from this distribution with magnitudes and redshifts updated by randomly shuffling about this error. Although these errors depend on a number of things including seeing, this method reproduces sensible results providing good agreement with existing high redshift clusters as shown in Figure \ref{fig:mockerror}.  

Generation of the artificial clusters now proceeds as follows. 

\begin{enumerate}
\item Randomly select a redshift and richness.
\item Select the closest redshift/richness bin.
\item Resample with replacement, a BCG and red sequence.
\item Apply a fixed $\Delta z$ to extrapolate from bin to mock cluster redshift.
\item K$+$e correct photometry
\item Sample suitable errors and reshuffle redshift and photometry.
\end{enumerate}

\noindent Finally, to simulate SDSS completeness levels members were removed with $i$-band $>21$ magnitudes.

Appropriate backgrounds for these artificial clusters are constructed by removing the red sequence from the original $10,000$ fields considered by GMPhoRCC. The list of backgrounds are binned in the same redshift/richness space as used previously according to the properties of the cluster. Real backgrounds are assigned by randomly selecting from the bin which best matches the properties of the artificial cluster. A total of $8745$ mocks were prepared with $0.05 < z < 1.1$ and $5 \leq n_{200} < 75$ by randomly inserting the artificial cluster within $3$ arc minutes of the centre of the assigned backgrounds. This method has the advantage of modelling the background as a function of the local neighbourhood where the background densities encounter would be typical for the given properties of the artificial cluster.

\subsubsection{Richness Consistency}

\begin{figure}
  \centering
  \includegraphics[width=240pt]{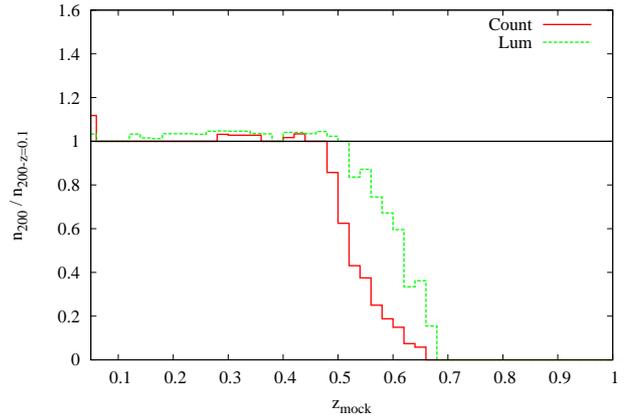} 
  \caption{A comparison of luminosity and counting $n_{200}$ as a function of redshift for mocks with an SDSS complete photometry cut, $i<21$ magnitudes. While incompleteness results in a loss of richness for $z>0.45$, at lower redshifts the GMPhoRCC richness is consistent. In addition the luminosity richness is shown to be able to reliably extrapolate beyond the $z\sim0.45$ completeness level up to $z\sim0.55$.}
  \label{fig:ngalsstockclcut}
\end{figure}

With the use of multiple red sequence bands it is necessary to ensure the GMPhoRCC richness estimate is consistent across the large redshift ranges considered which is indeed confirmed by analysis of the artificial clusters. Figure \ref{fig:ngalsstockclcut} shows how the GMPhoRCC estimate richness of artificial clusters, generated at $z=0.1$, evolves as these are extrapolated across $0.05 < z < 1.1$. While incompleteness results in a loss of richness above $z>0.45$, the GMPhoRCC estimate is consistent at lower redshifts. In addition the luminosity estimate has been shown to extrapolate into regions with incomplete photometry.

\subsection{Comparison with Mocks}

With the SDSS mocks, direct comparisons of GMPhoRCC estimates to `true' cluster values are possible. In order to asses accuracy and bias of GMPhoRCC the $i$-band$<21$ cut on mock members is not used, where the analysis of the full effect of incomplete photometry is deferred to a study of completeness in Section \ref{sec:comp}. 

\begin{figure*}
  \begin{minipage}{504pt}
    \includegraphics[width=240pt]{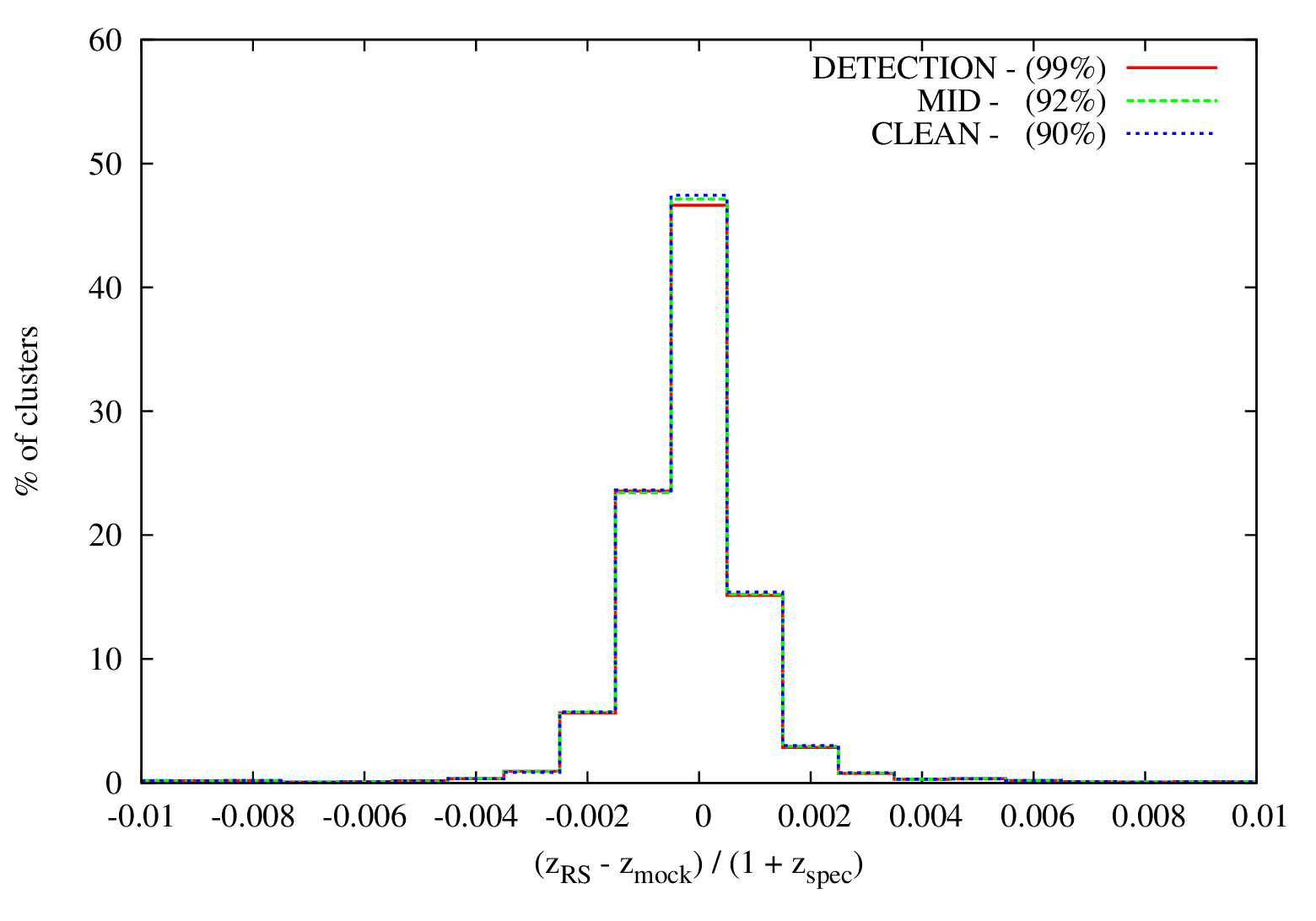}\hfill	
    \includegraphics[width=240pt]{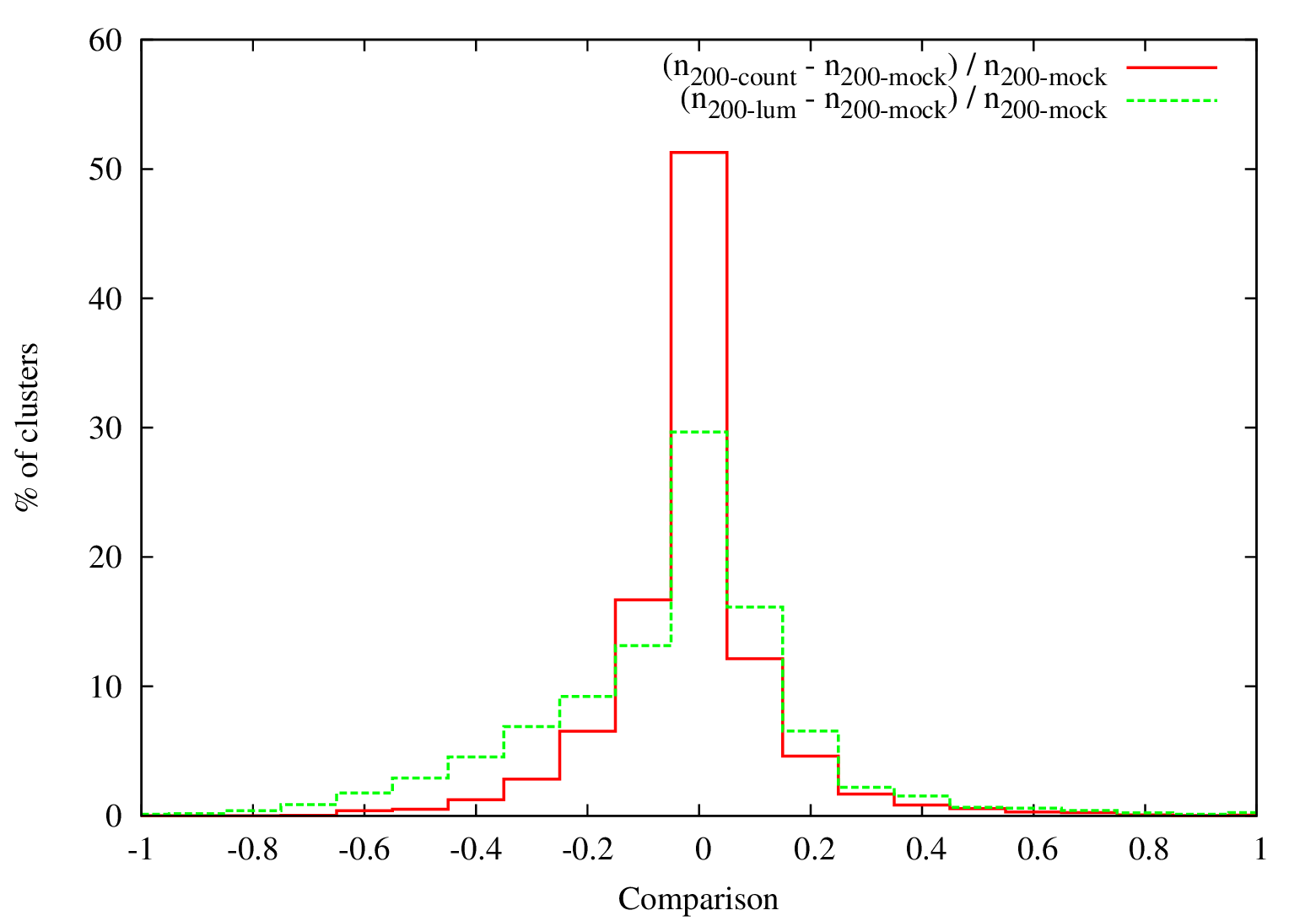} 
  \caption{Comparisons of GMPhoRCC estimates to actual values for $7050$ SDSS-like mock galaxy clusters. \textit{Left panel:} The distribution of the redshift comparisons where the results have been normalised and split into the separate quality subsets where the legend shows the fraction of the total clusters in each set. While the majority of all estimates are within $|z_{RS} - z_{mock}| / (1 + z_{mock}) <0.01$, the clean subset can again be seen to have removed the worst estimates with a greater fraction attaining this bound. This agrees well with results using real spectroscopic cluster showing accurate unbiased estimates. 
	\textit{Right panel:} The distribution of the $n_{200}$ comparison for the clean subset showing both the counting and luminosity function method. Both methods are observed to provide accurate and unbiased estimates of cluster richness. The larger scatters highlight the difficulty in recovering richness compared with redshift. The discreteness of $n_{200}$ and the sensitivity to discrepancies in redshift, $r_{200}$, $n_{gals}$, BCG identification, CMR modelling and projection effects, all contribute to the larger scatter. In addition the luminosity method is subject to the extra complexity and uncertainty introduced by fitting and integrating a luminosity function further increasing scatter.}
  \label{fig:mockcomp}
\end{minipage}
\end{figure*}

Of the $7050$ mock, estimates were found for $99.2$ percent and compared to the cluster values and shown in the left panel of Figure \ref{fig:mockcomp}. Redshift comparisons agree with those from real spectroscopic clusters where the GMPhoRCC are unbiased with the majority achieving $|z_{RS} - z_{mock}| / (1 + z_{mock}) <0.01$. In addition the clean subset attains a larger fraction within this bound and less contamination with outliers again highlighting the value of the quality control system. 

Richness comparisons, presented in right panel of Figure \ref{fig:mockcomp}, confirm that the GMPhoRCC estimate is unbiased with $|n_{200-count} - n_{200-mock}| = 0.01 \pm 0.005$ and $|n_{200-lum} - n_{200-mock}| = -0.03 \pm 0.02$. In addition it is clear both the counting and luminosity function method are able to adequately recover cluster richness. An accurate richness estimate is far more challenging to determine than redshift as evident by the larger scatter. These difficulties arise due to the discreteness of $n_{200}$ and the sensitivity to discrepancies in redshift, $r_{200}$, $n_{gals}$, BCG identification, CMR modelling and projection effects. In addition the luminosity method is subject to a lager scatter as a result of extra complexity and uncertainty introduced by fitting and integrating a luminosity function.

\subsection{Purity}

Although the target clusters for GMPhoRCC have already been detected in other wavebands (e.g. X-ray), it is important to understand purity when using the code to optically confirm a candidate or in cases where the candidate list may be contaminated. By using random real backgrounds only, purity is estimated as the fraction of fields where no cluster was detected i.e. detections in this case are impurities. While this only tackles the issue of false detections, the validity of the various GMPhoRCC estimates are assessed further in Section \ref{sec:comp}, which may be incorrect for a number of reason including projection effects.

Table \ref{table:pure} presents GMPhoRCC purity results which represent the probability that a candidate is in fact a cluster given that it was assigned a particular quality marker and richness. Very few spurious characterisations are found with high quality or richness, i.e. these have the highest probability of representing real clusters. Of particular note is the fact that candidates belonging to the clean subset, $q\geq3$, have a negligible probability of resulting from a false detection. It is noted that GMPhoRCC attains extremely high levels of purity compared with maxBCG which attains $\sim 93$ percent for clusters with $n_{200}=10$ and $\sim 99$ percent for $n_{200}=15$. Similarly compared with GMBCG which attains purity levels of $\sim 75$ percent for $n_{200}>10$ and $\sim 97$ percent for $n_{200}>25$.

\begin{table}
  \centering
  \caption[A list of the GMPhoRCC purity results based on richness and quality marker.]{A list of the GMPhoRCC purity results based on counting richness and quality marker. Very few spurious characterisations are found with high quality or richness, i.e. these candidates have the highest probability of representing real clusters. This analysis only concerns false detections where the validity of the richness estimate, which may be inaccurate due to projection effects, is further assessed in Section \ref{sec:comp}}
  \label{table:pure}
  \resizebox{240pt}{!}{
    \begin{tabular}{c c c c c}
    \hline\hline
     $q$ & $n_{200}>0$ &  $0 < n_{200} < 5 $ & $5 \leq n_{200} < 10 $ & $n_{200} \geq 10$\\
     \hline
    
    $>0$	& $79.7\%$  	  & $80.7\%$   	& $99.1\%$   	& $100.0\%$   \\	
    $1$ 	& $88.6\%$    	& $89.2\%$   	& $99.4\%$   	& $100.0\%$   \\
    $2$		& $92.0\%$      & $92.1\%$   	& $99.9\%$   	& $100.0\%$   \\
    $3$   & $99.2\%$  	  & $99.4\%$   	& $99.8\%$   	& $100.0\%$   \\
    
    \hline
  \end{tabular}
  }
\end{table}

\subsection{Completeness}
\label{sec:comp}

One of the most important properties to evaluate is completeness; this gives a measure of how well clusters are characterised across a range of redshifts and richnesses. Completeness is measured as the fraction of mock clusters where the estimated properties agree with the actual value within a given bound. In order to estimate the optical selection function, completeness is considered with respect to redshift, richness and BCG matching.

\subsubsection{Redshift Recovery}

\begin{figure*}
  \begin{minipage}{504pt}	
    \includegraphics[width=240pt]{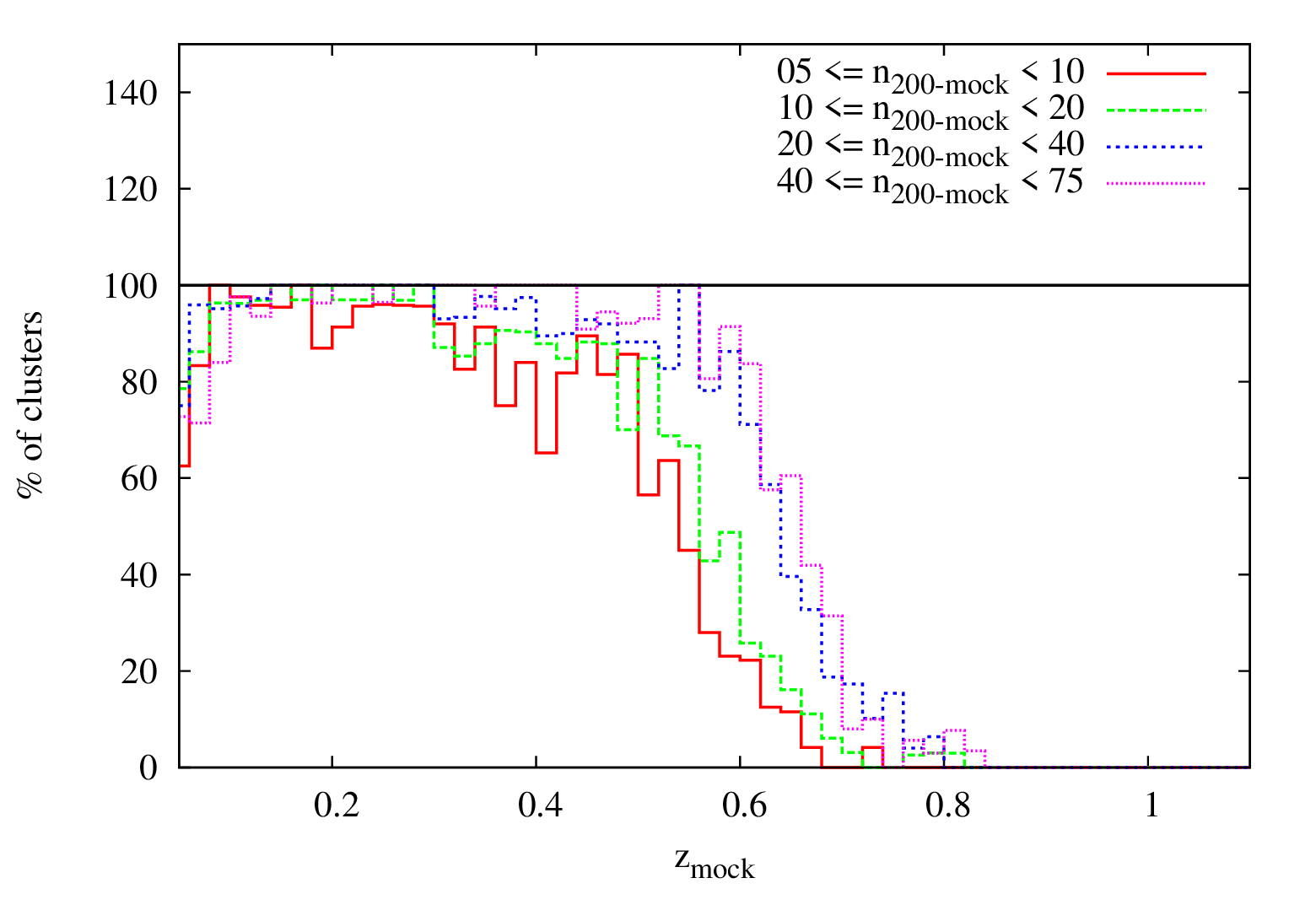}\hfill 
    \includegraphics[width=240pt]{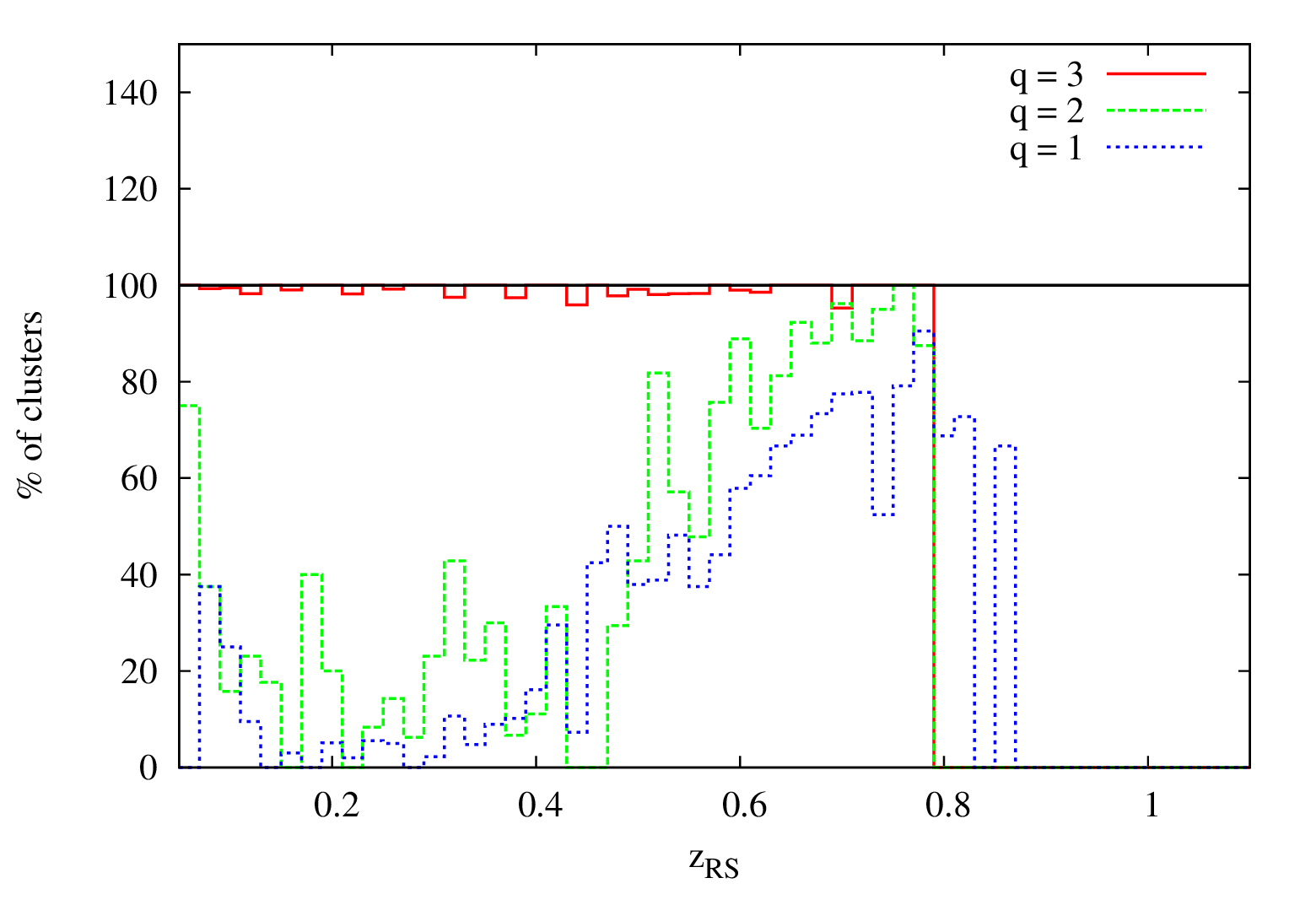} 
    \caption{\textit{Left panel:} The fraction of the clean subset of mock clusters for different richness bands where the redshift estimate is within $|z_{RS}-z_{mock}| < 0.03$. Low richness clusters are more sensitive to incomplete photometry due to the already low number of galaxies. Isolating the red sequence and estimating redshift is more challenging for low richness clusters than their high richness counterparts at the same redshift. Hence the ability to reliably estimate cluster redshift drops more quickly with redshift for groups than rich clusters. Due to this difficulty the clean set is also subject to an earlier reduction in completeness and a lower fraction of low richness clusters across all redshifts. 
		\textit{Right panel:} The fraction of mock clusters with a given $q$ which achieve the $|z_{RS}-z_{mock}| < 0.03$ bound. In addition to the clean subset, $q\geq3$, achieving a very high probability ($>97$ percent) that the redshift estimate is within $0.03$ of the mock value, those with lower quality for $z_{RS}<0.45$ have low probabilities ($<25$ percent), again showing the ability of the quality marker to identify and remove potential outliers. The sparse number of galaxies above $z>0.45$ in the SDSS DR10 and the mock background results in a low chance of spurious high redshift estimates, hence given $z_{RS}>0.45$ there is a larger probability the redshift is associated with the cluster. In addition, those with good high redshift estimates are more likely to be flagged as low richness due to the incomplete photometry. This leads to higher probabilities the estimate is associated with the cluster than expected for $q<3$.}
  \label{fig:mockcompletezn200allmaxBCG}
\end{minipage}
\end{figure*}

Using the fraction of the clean subset which attains the bound $|z_{rs}-z_{mock}| < 0.03$, comparable to typical SDSS photometric redshift errors, Figure \ref{fig:mockcompletezn200allmaxBCG} highlights completeness as a function of both richness and redshift with full results shown in Table \ref{table:zcomp}. For $z_{mock}<0.5$, the majority of the GMPhoRCC estimates are in very good agreement with the mock value, with high levels of completeness attained. Above this point photometry incompleteness results in difficulties in modelling the red sequence resulting in the lower completion. In addition to this, limitations in field area and poor contrast against the background for low redshift clusters, $z<0.1$, makes the red sequence more difficult to isolate and model, resulting in the lower fraction of clusters with good redshift estimates. As expected low richness cluster suffer from lower completeness due the difficulties in modelling sparse data sets. In addition these are seen to be more susceptible to photometry cuts resulting in the earlier reduction in completeness. 

\begin{table}
  \centering
  \caption[A summary of the redshift completeness results.]{A summary of the redshift completeness results.}
  \label{table:zcomp}
  \resizebox{240pt}{!}{
  \begin{tabular}{l c c c }
    \hline\hline
    Subset & Richness & Redshift & Completeness  \\
   
    \hline
                       & All                     & $0.05 < z < 0.65$ & $95.8\%$  \\
			                 & $5 \leq n_{200} < 10$   & $0.07 < z < 0.55$ & $97.7\%$  \\
			Detection  			 & $10 \leq n_{200} < 20$  & $0.07 < z < 0.55$ & $97.2\%$  \\
											 & $20 \leq n_{200} < 40$  & $0.05 < z < 0.65$ & $98.3\%$  \\
											 & $40 \leq n_{200} < 75$  & $0.05 < z < 0.65$ & $98.1\%$  \\
			\hline
			                 & All                     & $0.05 < z < 0.60$ & $92.6\%$  \\
			                 & $5 \leq n_{200} < 10$   & $0.07 < z < 0.55$ & $91.7\%$  \\
			Mid 						 & $10 \leq n_{200} < 20$  & $0.07 < z < 0.55$ & $93.0\%$  \\
											 & $20 \leq n_{200} < 40$  & $0.05 < z < 0.65$ & $93.8\%$  \\
											 & $40 \leq n_{200} < 75$  & $0.05 < z < 0.65$ & $94.8\%$  \\
			\hline
			                 & All                     & $0.05 < z < 0.60$ & $89.3\%$  \\
			                 & $5 \leq n_{200} < 10$   & $0.07 < z < 0.50$ & $89.4\%$  \\
			Clean						 & $10 \leq n_{200} < 20$  & $0.07 < z < 0.50$ & $92.2\%$  \\
											 & $20 \leq n_{200} < 40$  & $0.05 < z < 0.62$ & $92.6\%$  \\
											 & $40 \leq n_{200} < 75$  & $0.05 < z < 0.62$ & $93.6\%$  \\
    \hline

  \end{tabular}}
\end{table}

Extending this completeness analysis by considering the subset of clusters with a given $q$ and $z_{RS}$, the accuracy of the GMPhoRCC redshift is estimated. Shown in the right panel of Figure \ref{fig:mockcompletezn200allmaxBCG} is the fraction of these subsets attaining the redshift bound which represents the probability that given a cluster has a specific $q$ and $z_{rs}$ achieves $|z_{rs}-z_{mock}| < 0.03$. In addition to the clean subset, $q\geq3$, achieving a very high probability ($>97$ percent) that the redshift estimate is within $0.03$ of the mock value, those with lower quality for $z<0.45$ have low probabilities ($<25$ percent), again showing the ability of the quality subsets to identify and remove potential outliers. The sparse number of galaxies above $z>0.45$ in the SDSS DR10 and the mock background results in a low chance of spurious high redshift estimates, hence given $z_{RS}>0.45$ there is a larger probability the redshift is associated with the cluster. In addition, those with good high redshift estimates are more likely to be flagged as low richness due to the incomplete photometry. This leads to higher probabilities the estimate is associated with the cluster than expected for $q<3$. While adjustments could be made to the quality subsets to take advantage of this increased probability it is noted that for $z>0.45$ the lower quality mainly result from low numbers of galaxies due to incompleteness and hence the current quality subsets are necessary to maintain cleanliness for both redshift and richness estimates. A full set of probabilities for each quality marker and several bounds are presented in Table \ref{table:zcompprob}.

\begin{table}
  \centering
  \caption[A list of the probabilities that a redshift estimate is within various bounds of the actual value given the $z_{RS}$ estimate and $q$ value of the cluster.]{A list of the probabilities that a redshift estimate is within various bounds of the actual value given the $z_{RS}$ estimate and $q$ value of the cluster, where $\Delta z = |z_{RS} - z_{mock}|$. The increase in probability for low quality high redshift clusters is clear. The sparse number of galaxies above $z>0.45$ in the SDSS DR10 and the mock background results in a low chance of spurious high redshift estimates hence given $z_{RS}>0.45$ there is a larger probability the redshift is associated with the cluster than expected for those with $q<3$. }
  \label{table:zcompprob} 
  \resizebox{240pt}{!}{
    \begin{tabular}{c c c c c }
      \hline\hline
      $q$ & Redshift Range & p($\Delta z < 0.01$) & p($\Delta z < 0.03$) &p($\Delta z < 0.05$)  \\
      \hline

      1  			& $0.05 < z_{RS} < 0.50$ & $0.11$ & $0.11$& $0.12$ \\			
			2				& $0.05 < z_{RS} < 0.50$ & $0.16$ & $0.19$& $0.25$ \\
			3				& $0.05 < z_{RS} < 0.80$ & $0.96$ & $0.99$& $0.99$ \\
      \hline
      1 			& $0.50 < z_{RS} < 0.80$ & $0.45$ & $0.61$& $0.66$ \\
			2 			& $0.50 < z_{RS} < 0.80$ & $0.68$ & $0.81$& $0.86$ \\
      \hline
    \end{tabular}
  }
\end{table}

\subsubsection{Richness Recovery}

While two richnesses are investigated by GMPhoRCC, $n_{200}$ best represents an optical mass proxy, considering galaxies within a characteristic radius, rather than the fixed aperture of $n_{gals}$, and hence is the subject of this section. With the extra $n_{gals}$ step additional sources of error are introduced and with $n_{200}$ highly sensitive to correct CMR modelling, BCG selection and redshift richness attains much a much larger spread about the mock value and thus relatively large completeness bounds are considered.

\begin{figure*}
  \begin{minipage}{504pt}
    \includegraphics[width=240pt]{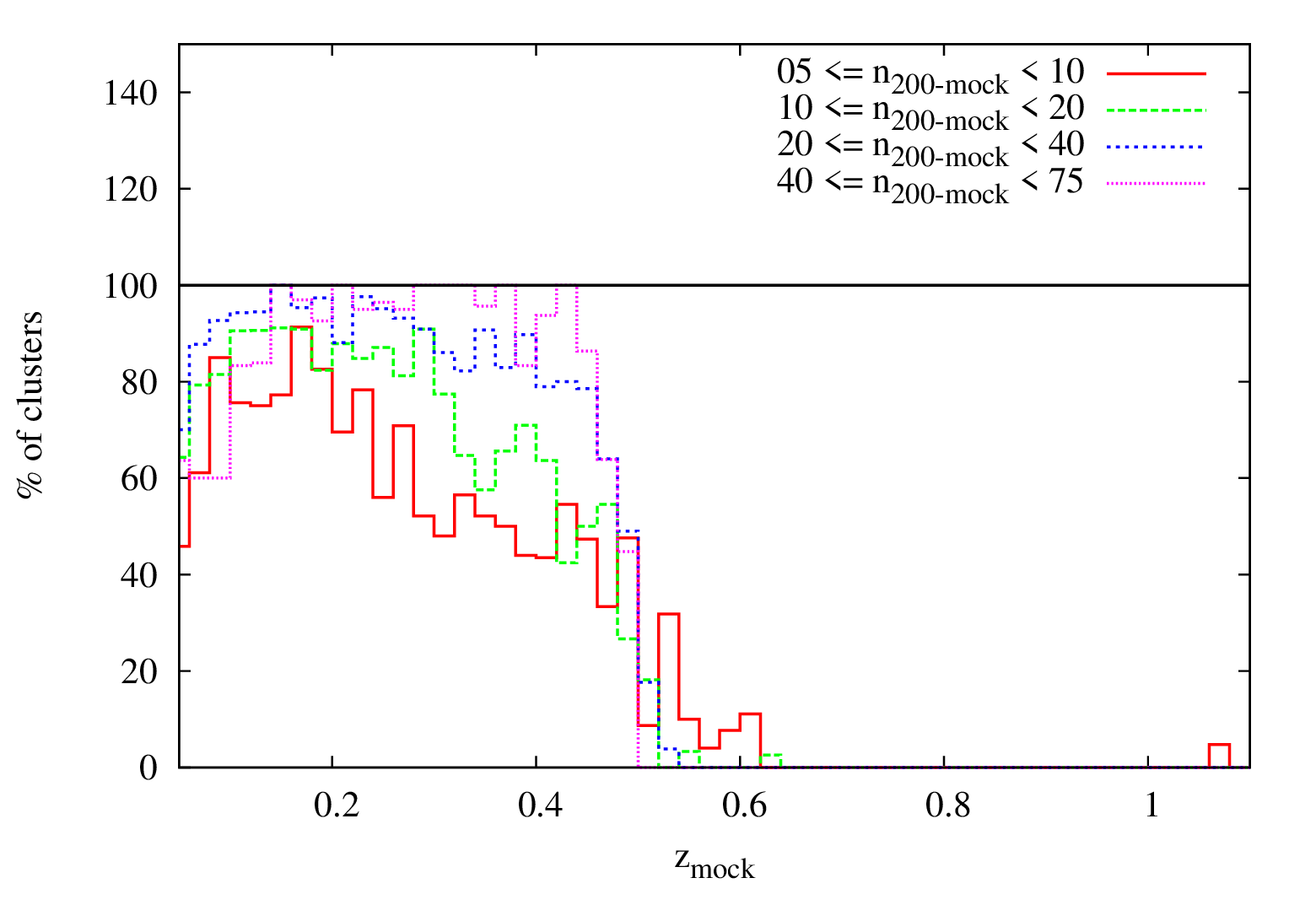}\hfill 
    \includegraphics[width=240pt]{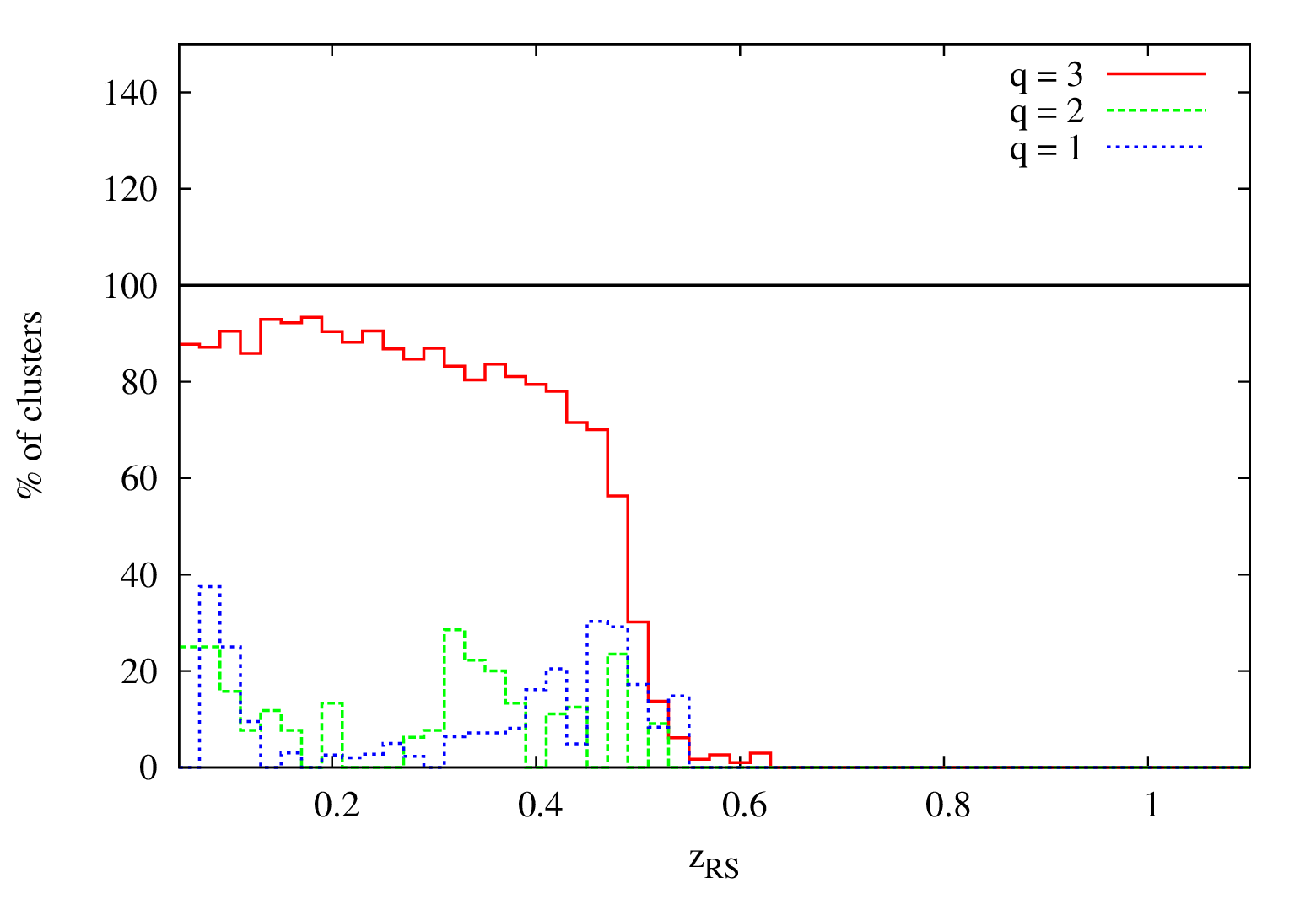} 
    \caption{\textit{Left panel:} The fraction of the clean subset of mock clusters in richness bands where the $n_{200-count}$ estimate was within $25$ percent of the actual value. Above $z>0.45$ incompleteness in photometry results in the sharp decline in richness recovery rates. In addition low richness clusters are more susceptible to this decline due to modelling the already low number of galaxies. In addition to this, $n_{200}$ is highly sensitive to errors in redshift, BCG selection and $n_{gals}$ resulting in lower completion rates than with redshift.
      \textit{Right panel:} The fraction of mock clusters with a given $q$ where the $n_{200}$ estimate was within $25$ percent of the original value. In addition to the clean subset, $q\geq3$, achieving a high probability ($>80$ percent) that the richness estimate is within $25$ percent of the mock value, those with lower quality have very low probabilities ($<15$ percent), again showing the ability of the quality subsets to identify and remove potential outliers.}
    \label{fig:mockcompletengcn200detectionmaxBCG}
  \end{minipage}
\end{figure*}

Figure \ref{fig:mockcompletengcn200detectionmaxBCG} highlights richness completeness as the fraction of the clean subset where the counting richness is within $25$ percent of the mock value. Completeness in both the counting and luminosity estimate tails off above $z>0.45$ due to incomplete photometry where cluster galaxies become too faint for reliable detection. It is noted that the luminosity method is able to extrapolate richness resulting in a slower reduction with redshift and higher completeness than the counting method for $z>0.45$. 

As stated in the previous sections, low richness clusters are difficult to model and analyse due to difficulties in fitting distributions to a small number of galaxies, and this is reflected in the lower completeness rates. In addition to this, background fluctuations which, in this case can cause as much as an $80\%$ discrepancy due to discreteness further reduces completeness. Table \ref{table:richcomp} summarises and extends these results to the luminosity richness, $n_{200-lum}$.

\begin{table}
  \centering
  \caption[A summary of the completeness results for $n_{200}$]{A summary of results where the top and bottom Tables present completeness where the $n_{200-count}$ and $n_{200-lum}$ estimates respectively are within $25$ percent of the mock value. Due to the extrapolation, the luminosity method attains a deeper redshift range than counting before completeness rates decline.}
  \label{table:richcomp}
  (a) The counting richness, $n_{200-count}$\\
  \vspace{1mm}
  \resizebox{240pt}{!}{
    \begin{tabular}{l c c c}
      
      \hline\hline
      Subset & Richness & Redshift & Completeness\\
      \hline
                       & All                     & $0.07 < z < 0.45$ & $85.0\%$  \\
			                 & $5  \leq n_{200} < 10$  & $0.07 < z < 0.45$ & $67.5\%$  \\
			Detection  			 & $10 \leq n_{200} < 20$  & $0.07 < z < 0.45$ & $81.2\%$  \\
											 & $20 \leq n_{200} < 40$  & $0.07 < z < 0.45$ & $92.8\%$  \\
											 & $40 \leq n_{200} < 75$  & $0.07 < z < 0.45$ & $93.2\%$  \\
			\hline
			                 & All                     & $0.07 < z < 0.45$ & $83.3\%$  \\
			                 & $5  \leq n_{200} < 10$  & $0.07 < z < 0.45$ & $65.6\%$  \\
			Mid 						 & $10 \leq n_{200} < 20$  & $0.07 < z < 0.45$ & $79.2\%$  \\
											 & $20 \leq n_{200} < 40$  & $0.07 < z < 0.45$ & $91.0\%$  \\
											 & $40 \leq n_{200} < 75$  & $0.07 < z < 0.45$ & $92.3\%$  \\
			\hline
			                 & All                     & $0.07 < z < 0.45$ & $82.4\%$  \\
			                 & $5  \leq n_{200} < 10$  & $0.07 < z < 0.45$ & $64.3\%$  \\
			Clean						 & $10 \leq n_{200} < 20$  & $0.07 < z < 0.45$ & $77.9\%$  \\
											 & $20 \leq n_{200} < 40$  & $0.07 < z < 0.45$ & $90.7\%$  \\
											 & $40 \leq n_{200} < 75$  & $0.07 < z < 0.45$ & $91.3\%$  \\
     
        \hline
      \end{tabular}
  }\\
  \vspace{3mm}
  (b) The luminosity richness, $n_{200-lum}$\\
  \vspace{1mm}
  \resizebox{240pt}{!}{
      \begin{tabular}{l c c c }
        \hline\hline
        Subset & Richness & Redshift & Completeness  \\
        \hline
                       & All                     & $0.07 < z < 0.50$ & $81.8\%$  \\
			                 & $5  \leq n_{200} < 10$  & $0.07 < z < 0.45$ & $67.5\%$  \\
			Detection  			 & $10 \leq n_{200} < 20$  & $0.07 < z < 0.45$ & $81.2\%$  \\
											 & $20 \leq n_{200} < 40$  & $0.07 < z < 0.52$ & $87.5\%$  \\
											 & $40 \leq n_{200} < 75$  & $0.07 < z < 0.52$ & $88.0\%$  \\
			\hline
			                 & All                     & $0.07 < z < 0.50$ & $79.5\%$  \\
			                 & $5  \leq n_{200} < 10$  & $0.07 < z < 0.45$ & $65.6\%$  \\
			Mid 						 & $10 \leq n_{200} < 20$  & $0.07 < z < 0.45$ & $79.2\%$  \\
											 & $20 \leq n_{200} < 40$  & $0.07 < z < 0.52$ & $85.1\%$  \\
											 & $40 \leq n_{200} < 75$  & $0.07 < z < 0.52$ & $86.9\%$  \\
			\hline
			                 & All                     & $0.07 < z < 0.50$ & $78.6\%$  \\
			                 & $5  \leq n_{200} < 10$  & $0.07 < z < 0.45$ & $64.3\%$  \\
			Clean						 & $10 \leq n_{200} < 20$  & $0.07 < z < 0.45$ & $77.9\%$  \\
											 & $20 \leq n_{200} < 40$  & $0.07 < z < 0.52$ & $84.8\%$  \\
											 & $40 \leq n_{200} < 75$  & $0.07 < z < 0.52$ & $86.0\%$  \\
		
        \hline
      \end{tabular}
    
  }
\end{table}

Again completeness is considered with respect to subsets with a given $q$ and $z_{rs}$ to estimate the accuracy of GMPhoRCC richness. Shown in the right panel of Figure \ref{fig:mockcompletengcn200detectionmaxBCG} is the fraction of these subsets which attain the richness bound, representing the probability that given a cluster has a specific $q$ and $z_{rs}$ that the richness estimate is within $25$ percent of the mock value. A full set of probabilities for each quality marker and several bounds are presented in Table \ref{table:richcompprob}.  In addition to the clean subset, $q\geq3$, achieving a very high probability ($>80$ percent) that the richness estimate is within $25$ percent of the mock value, those with higher quality markers have low probabilities ($<15$ percent), again showing the ability of the quality subsets to identify and remove potential outliers. In addition, while the counting richness has negligible probability of matching the mock at high redshift ($0.5 < z_{RS} < 0.8$), the luminosity method is clearly able to extrapolate, achieving a $30$ percent probability that the richness is with 25 percent of the mock value. 

\begin{table}
  \centering
  \caption[A list of the probabilities that the $n_{200}$ estimate is within various bounds of the actual value given the $z_{RS}$ estimate and the quality marker of the cluster.]{A list of the probabilities that the $n_{200}$ estimate is within various bounds of the actual value given the $z_{RS}$ estimate and the quality marker of the cluster. The top and bottom Tables present results for the counting and luminosity richnesses respectively, where $\Delta n_{c} = |n_{200-count} - n_{200-mock}| / n_{200-mock}$ and $\Delta n_{l} = |n_{200-lum} - n_{200-mock}| / n_{200-mock}$. In addition to the clean subset, $q\geq3$, achieving the highest probabilities that the richness estimate is within given bounds of the mock value, those with lower quality have very low probabilities ($<15$ percent), again showing the ability of the quality subsets to identify and remove potential outliers. In addition, while the counting richness has negligible probability of matching the mock at high redshift ($0.5 < z_{RS} < 0.8$), the luminosity method is clearly able to extrapolate, achieving a $30$ percent probability that the richness is with 25 percent of the mock value. }
  \label{table:richcompprob}
  (a) The counting richness $n_{200-count}$\\
  \vspace{1mm}
  \resizebox{240pt}{!}{
      \begin{tabular}{c c c c c }
	\hline\hline
	$q$ & Redshift Range & {p($\Delta n_{c} < 0.1$)} & {p($\Delta n_{c} < 0.25$)} & {p($\Delta n_{c} < 0.5$)} \\
	\hline
	
	    1  			& $0.05 < z_{RS} < 0.45$ & $0.03$ & $0.06$& $0.08$ \\			
			2 			& $0.05 < z_{RS} < 0.45$ & $0.05$ & $0.11$& $0.20$ \\
			3				& $0.05 < z_{RS} < 0.45$ & $0.53$ & $0.86$& $0.97$ \\
		
	\hline
      \end{tabular}
  }\\
  \vspace{3mm}
  (b) The luminosity richness $n_{200-lum}$\\
  \vspace{1mm}
  \resizebox{240pt}{!}{ 
      \begin{tabular}{c c c c c }
	\hline\hline
	$q$ & Redshift Range & {p($\Delta n_{l} < 0.1$)} & {p($\Delta n_{l} < 0.25$)} &{p($\Delta n_{l} < 0.5$)}  \\
	\hline
	
	1  			    & $0.05 < z_{RS} < 0.50$ & $0.04$ & $0.09$& $0.12$ \\			
	2 					& $0.05 < z_{RS} < 0.50$ & $0.05$ & $0.11$& $0.20$ \\
	3						& $0.05 < z_{RS} < 0.50$ & $0.51$ & $0.83$& $0.96$ \\
	\hline
	1  			    & $0.50 < z_{RS} < 0.80$ & $0.04$ & $0.09$& $0.22$ \\			
	2 					& $0.50 < z_{RS} < 0.80$ & $0.04$ & $0.11$& $0.24$ \\
	3						& $0.50 < z_{RS} < 0.80$ & $0.13$ & $0.30$& $0.58$ \\

	\hline
      \end{tabular}
  } 
\end{table}

\subsubsection{BCG Identification}

Identifying the correct BCG is not only hugely important for subsequent cosmology but also for calculating cluster richness. This analysis considers two scenarios, one where the BCG is correctly identified and one where any cluster member is selected as the BCG. While correctly matching the BCG shows the strongest evidence GMPhoRCC has suitably modelled the red sequence, even matching to a cluster member suggests the CMR is a reasonable representation of the cluster. 

\begin{figure*}
  \begin{minipage}{504pt}	
    \includegraphics[width=240pt]{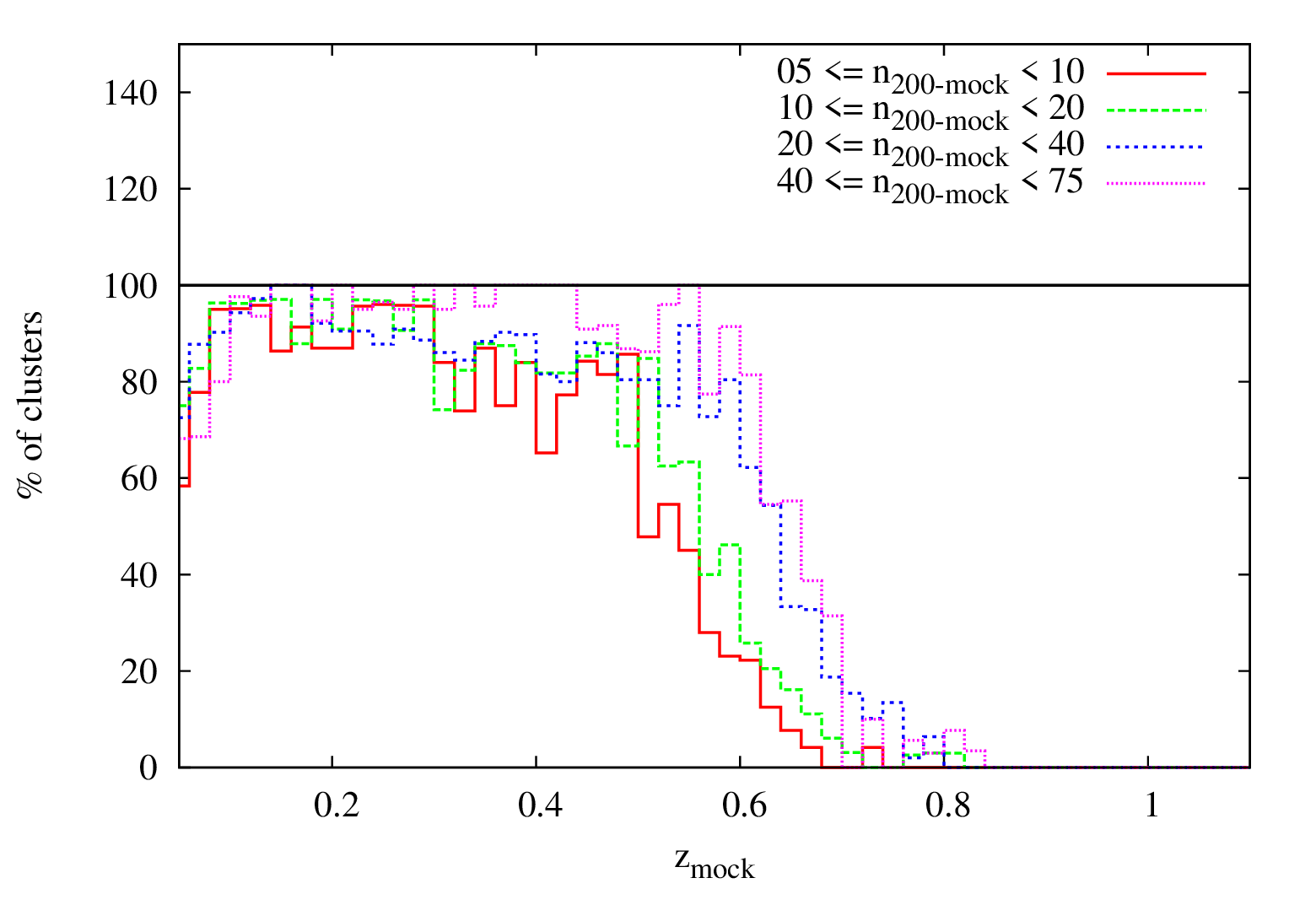}\hfill 
    \includegraphics[width=240pt]{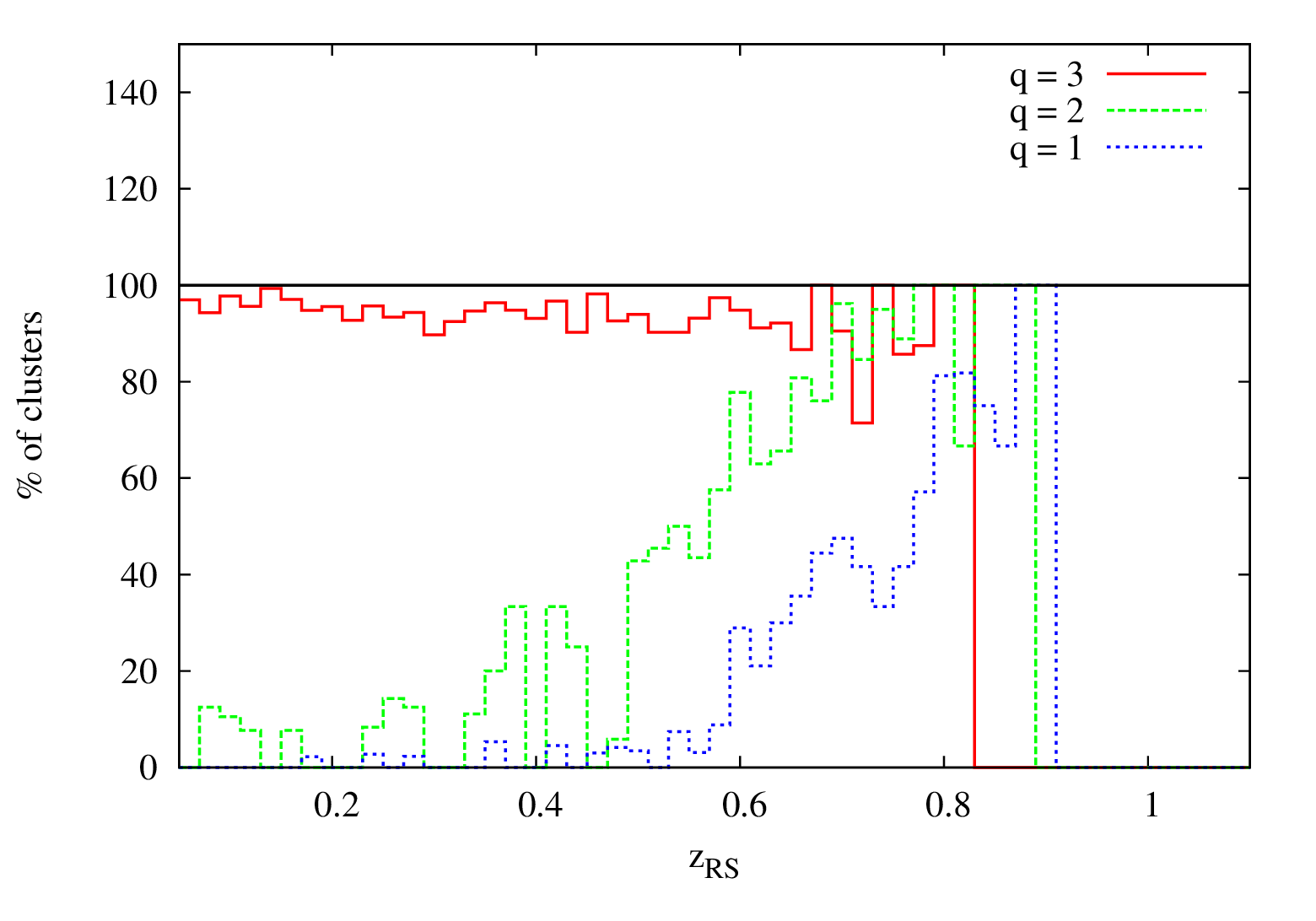} 
    \caption{\textit{Left panel:} The fraction of the clean subset of mock clusters for different richness bands where the BCG has been correctly identified. Above $z>0.5$ incompleteness in photometry results in difficulties fitting the red sequence resulting the reduction of matching rates. In addition with the difficulty in modelling low richness clusters these suffer from earlier declines and lower matching rates.
      \textit{Right panel:} The fraction of mock clusters with a given $q$ where the BCG has been correctly identified. In addition to the clean subset, $q\geq3$, achieving a very high probability ($>90$ percent) that the BCG has been correctly identified, those with lower quality for $z<0.45$ have low probabilities ($<10$ percent), again showing the ability of the quality subsets to identify and remove cases where the red sequence has not been well modelled. Again the sparse number of galaxies above $z>0.45$ in the SDSS DR10 and the mock background results in a low chance of spurious high redshift estimates hence given $z_{RS}>0.45$ there is a larger probability the CMR and BCG are associated with the cluster. In addition, those with suitable CMRs at high redshifts are more likely to be flagged as low richness due to the incomplete photometry. This leads higher probabilities the CMR and BCG is associated with the cluster than expected for $q<3$.}
    \label{fig:mockcompletebcgn200allmaxBCG}
  \end{minipage}
\end{figure*}

Mismatching the BCG results from two main issues, background interlopers and poor red sequence modelling. While mismatching to a background galaxy is easier to find with the quality markers due to inconsistencies in redshift, matching to another cluster member can be more challenging to identify. Figure \ref{fig:mockcompletebcgn200allmaxBCG} shows the fraction of the clean subset of mocks where the BCG has been correctly matched.  As photometry becomes incomplete issues with fitting the red sequence due to the lower number of galaxies gives rise to the lower fraction matched above $z>0.5$. In addition to this the difficulty in modelling the red sequence at low redshift, $z<0.1$, due to poor contrast against the background and limitations in the field area, result in a smaller fraction of these mocks with correctly matched BCG.

Again it is expected that a smaller fraction of low richness clusters have suitably determined CMRs due to the difficulty in modelling a sparse number of galaxies and this is reflected in the lower BCG match rates. Table \ref{table:bcgcomp} summarises and extends these results to correct BCG matching and cluster member BCG matching for each of the quality subsets.

Again considering the BCG matching fractions with respect to subsets with a given $q$ and $z_{RS}$ gives an estimate of the probability that the BCG has been correctly matched given the cluster is consistent with the subset. The right panel of Figure \ref{fig:mockcompletebcgn200allmaxBCG} shows these fractions for clusters where the BCG has been correctly identified. A full set of probabilities for each quality marker and the different BCG sources are presented in Table \ref{table:bcgcompprob}. In addition to the clean subset, $q\geq3$, achieving a very high probability ($>90$ percent) that the BCG has been correctly identified, those with lower quality for $z<0.45$ have low probabilities ($<10$ percent), again showing the ability of the quality subsets to identify and remove cases where the red sequence has not been well modelled. The sparse number of galaxies above $z>0.45$ in the SDSS DR10 and the mock background results in a low chance of spurious high redshift estimates hence given $z_{RS}>0.45$ there is a larger probability the CMR and BCG are associated with the cluster. In addition, those with suitable CMRs at high redshifts are more likely to be flagged as low richness due to the incomplete photometry. This leads to higher probabilities the CMR and BCG is associated with the cluster than expected for $q<3$. Again no adjustments are made to the quality subsets since incomplete photometry becomes an issue for $z_{RS}>0.45$ with the current subsets necessary to maintain cleanliness for both redshift and richness estimates. 

\begin{table}
  \centering
  \caption[A summary of the completeness results for various sources of the BCG]{A summary of the completeness results where the top Table presents completeness where the BCG has been correctly identified and the bottom where the BCG has been matched to any cluster member. A smaller fraction of low richness clusters have suitable CMRs due to the difficulty in modelling a sparse number of galaxies resulting in lower completeness rates in both cases. }
  \label{table:bcgcomp}
   (a) Correctly identified BCG\\
   \vspace{1mm}
  \resizebox{240pt}{!}{
   
      \begin{tabular}{l c c c}
        \hline\hline
        Subset & Richness & Redshift & Completeness\\
        \hline
                          & All                     & $0.05 < z < 0.60$ & $87.9\%$  \\
			                 & $5 \leq n_{200} < 10$   & $0.05 < z < 0.55$ & $86.6\%$  \\
			Detection  			 & $10 \leq n_{200} < 20$  & $0.05 < z < 0.55$ & $89.1\%$  \\
											 & $20 \leq n_{200} < 40$  & $0.05 < z < 0.62$ & $87.2\%$  \\
											 & $40 \leq n_{200} < 75$  & $0.05 < z < 0.62$ & $93.1\%$  \\
			\hline
			                 & All                     & $0.05 < z < 0.58$ & $87.8\%$  \\
			                 & $5 \leq n_{200} < 10$   & $0.05 < z < 0.50$ & $87.3\%$  \\
			Mid 						 & $10 \leq n_{200} < 20$  & $0.05 < z < 0.55$ & $88.7\%$  \\
											 & $20 \leq n_{200} < 40$  & $0.05 < z < 0.62$ & $86.9\%$  \\
											 & $40 \leq n_{200} < 75$  & $0.05 < z < 0.62$ & $92.6\%$  \\
			\hline
			                 & All                     & $0.05 < z < 0.58$ & $86.1\%$  \\
			                 & $5 \leq n_{200} < 10$   & $0.07 < z < 0.50$ & $86.6\%$  \\
			Clean						 & $10 \leq n_{200} < 20$  & $0.05 < z < 0.55$ & $87.1\%$  \\
											 & $20 \leq n_{200} < 40$  & $0.05 < z < 0.62$ & $85.9\%$  \\
											 & $40 \leq n_{200} < 75$  & $0.05 < z < 0.62$ & $91.7\%$  \\
        \hline
      \end{tabular}
  }\\
  \vspace{3mm}
  (b) BCG matched to any cluster member\\
  \vspace{1mm}
  \resizebox{240pt}{!}{
    \begin{tabular}{l c c c }
	\hline\hline
	Subset & Richness & Redshift & Completeness  \\
	\hline
	                     & All                     & $0.05 < z < 0.60$ & $91.2\%$  \\
			                 & $5 \leq n_{200} < 10$   & $0.05 < z < 0.55$ & $88.0\%$  \\
			Detection  			 & $10 \leq n_{200} < 20$  & $0.05 < z < 0.55$ & $90.7\%$  \\
											 & $20 \leq n_{200} < 40$  & $0.05 < z < 0.62$ & $92.6\%$  \\
											 & $40 \leq n_{200} < 75$  & $0.05 < z < 0.62$ & $96.7\%$  \\
			\hline
			                 & All                     & $0.05 < z < 0.58$ & $90.9\%$  \\
			                 & $5 \leq n_{200} < 10$   & $0.05 < z < 0.50$ & $88.8\%$  \\
			Mid 						 & $10 \leq n_{200} < 20$  & $0.05 < z < 0.55$ & $90.2\%$  \\
											 & $20 \leq n_{200} < 40$  & $0.05 < z < 0.62$ & $92.2\%$  \\
											 & $40 \leq n_{200} < 75$  & $0.05 < z < 0.62$ & $95.4\%$  \\
			\hline
			                 & All                     & $0.05 < z < 0.58$ & $88.7\%$  \\
			                 & $5 \leq n_{200} < 10$   & $0.05 < z < 0.50$ & $87.5\%$  \\
			Clean						 & $10 \leq n_{200} < 20$  & $0.05 < z < 0.55$ & $88.3\%$  \\
											 & $20 \leq n_{200} < 40$  & $0.07 < z < 0.62$ & $91.1\%$  \\
											 & $40 \leq n_{200} < 75$  & $0.05 < z < 0.62$ & $93.1\%$  \\
	\hline
      \end{tabular}
  }      
\end{table}

\begin{table}
  \centering
  \caption[A list of the probabilities that the BCG has been identified from various sources given the cluster has a specific quality marker and redshift estimate.]{A list of the probabilities that the BCG has been matched to various sources given the cluster has a specific quality marker and redshift estimate. In both scenarios the clean set demonstrates the highest probability that the red sequence has been suitably modelled.}
  \label{table:bcgcompprob}
  \resizebox{240pt}{!}{
  
      \begin{tabular}{c c c c}
	\hline\hline
	$q$ & Redshift Range & p(Correct BCG) & p(Cluster Member BCG)  \\
	\hline
	
	1  		& $0.05 < z_{RS} < 0.60$ & $0.03$  & $0.03$  \\			
	2 		& $0.05 < z_{RS} < 0.50$ & $0.10$  & $0.16$  \\
	3			& $0.05 < z_{RS} < 0.80$ & $0.94$  & $0.97$  \\
	\hline
	1  		& $0.60 < z_{RS} < 0.80$ & $0.39$  & $0.39$  \\			
	2 		& $0.50 < z_{RS} < 0.80$ & $0.72$ & $0.75$ \\
	\hline
      \end{tabular}
  } 
\end{table}

\section{Discussion and Conclusions}
\label{sec:concl}

Presented in this paper is the \textbf{G}aussian \textbf{M}ixture full \textbf{Ph}otometric \textbf{R}ed sequence \textbf{C}luster \textbf{C}haracteriser (GMPhoRCC), which is designed to take cluster candidates,  previously detected, and provide an optical confirmation and characterisation based on the red sequence. GMPhoRCC has been designed specifically to attain estimates of redshift, richness and the red sequence CMR and offers many advantages over existing algorithms including, treatment of multi-modal distributions, treatment of a variable width full CMR red sequence, richness extrapolation and quality control. One of the most important features developed is the flag and quality control procedure. By flagging issues, particularly low richness and inconsistent red sequence and BCG redshifts, potential catastrophic failures can be identified and removed from cleaner subsets.

Comparisons with other characterisation methods highlights the advantages of GMPhoRCC. Using a sample of $4501$ clusters taken from the GMBCG \citep{gmbcg}, NORAS \citep{noras}, REFLEX \citep{reflex} and XCS \citep{xcsoptical} catalogues, GMPhoRCC redshift estimates are compared to spectra showing low scatter ($\sigma_{\delta z / (1+z)} \sim 0.026$) around the actual value. In addition applying the quality control to produce a clean subset removes most outliers giving a much tighter agreement, $\sigma_{\delta z / (1+z)} \sim 0.017$ showing significant improvement over maxBCG, $\sigma_{\delta z / (1+z)} \sim 0.025$, and XCS, $\sigma_{\delta z / (1+z)} \sim 0.050$. The high accuracy of GMPhoRCC is also demonstrated with a significant percentage ($\sim75\%$) of all redshift estimates from the clean subset agreeing within $|z_{RS}-z_{spec}|<0.01$.

While analysing known clusters provides useful feedback, comparisons with those with known properties are far more valuable, hence the remaining evaluation of GMPhoRCC proceeded with the use of empirical mock galaxy clusters. These mocks were produced by stacking red sequence galaxies from existing clusters, analysed using data from the Sloan Digital Sky Survey (SDSS), in redshift-richness bins from which new sequences are resampled. This extends the similar approach of maxBCG and GMBCG where only rich clusters are used as seeds to generate mocks with a range of properties. 

Assessment of the optical selection function proceeded with the consideration of completeness, the fraction of mocks with characterisations within given bounds of the actual value. First incomplete photometry, simulated by an $i$-band $< 21$ cut, is shown to remove members for clusters with $z>0.45$. Redshift completeness, the fraction of clusters within $0.03$ of the mock value, is not immediately hindered by the photometry attaining $93\%$ for $0.05 < z <0.62$ for clusters with a richness greater than $20$. With the large scatters in the estimates, richness attains lower completeness rates, mostly due to projection effects and background fluctuations as also noted by \cite{gmbcg}. The fraction of clusters within $25\%$ of the mock value, defining completeness, is measured as $91\%$ for $0.07 < z < 0.45$ for clusters with richness greater than $20$, $78\%$ for those with richness between $10$ and $20$, and $64 \%$ for those with richnesses less than $10$.

Additionally evaluation with mocks had confirmed the value of the quality control system showing a high probability that given a cluster is in the clean set that the redshift and richness estimates are within a given bound of the mock value. Most importantly it was shown that those with lower quality markers, indicating less confidence in the characterisation, show much smaller probabilities confirming that the quality control is effective in identifying potential catastrophic failures.

\section*{Acknowledgements}

This paper makes use of optical data from the Sloan Digital Sky Survey, SDSS-III. Funding for SDSS-III has been provided by the Alfred P. Sloan Foundation, the Participating Institutions, the National Science Foundation, and the U.S. Department of Energy Office of Science. The SDSS-III web site is \url{http://www.sdss3.org/.}

SDSS-III is managed by the Astrophysical Research Consortium for the Participating Institutions of the SDSS-III Collaboration including the University of Arizona, the Brazilian Participation Group, Brookhaven National Laboratory, Carnegie Mellon University, University of Florida, the French Participation Group, the German Participation Group, Harvard University, the Instituto de Astrofisica de Canarias, the Michigan State/Notre Dame/JINA Participation Group, Johns Hopkins University, Lawrence Berkeley National Laboratory, Max Planck Institute for Astrophysics, Max Planck Institute for Extraterrestrial Physics, New Mexico State University, New York University, Ohio State University, Pennsylvania State University, University of Portsmouth, Princeton University, the Spanish Participation Group, University of Tokyo, University of Utah, Vanderbilt University, University of Virginia, University of Washington, and Yale University.




\bibliographystyle{mnras}
\bibliography{HoodMann20161123} 

\begin{thebibliography}{}
\makeatletter
\relax
\def\mn@urlcharsother{\let\do\@makeother \do\$\do\&\do\#\do\^\do\_\do\%\do\~}
\def\mn@doi{\begingroup\mn@urlcharsother \@ifnextchar [ {\mn@doi@}
  {\mn@doi@[]}}
\def\mn@doi@[#1]#2{\def\@tempa{#1}\ifx\@tempa\@empty \href
  {http://dx.doi.org/#2} {doi:#2}\else \href {http://dx.doi.org/#2} {#1}\fi
  \endgroup}
\def\mn@eprint#1#2{\mn@eprint@#1:#2::\@nil}
\def\mn@eprint@arXiv#1{\href {http://arxiv.org/abs/#1} {{\tt arXiv:#1}}}
\def\mn@eprint@dblp#1{\href {http://dblp.uni-trier.de/rec/bibtex/#1.xml}
  {dblp:#1}}
\def\mn@eprint@#1:#2:#3:#4\@nil{\def\@tempa {#1}\def\@tempb {#2}\def\@tempc
  {#3}\ifx \@tempc \@empty \let \@tempc \@tempb \let \@tempb \@tempa \fi \ifx
  \@tempb \@empty \def\@tempb {arXiv}\fi \@ifundefined
  {mn@eprint@\@tempb}{\@tempb:\@tempc}{\expandafter \expandafter \csname
  mn@eprint@\@tempb\endcsname \expandafter{\@tempc}}}

\bibitem[\protect\citeauthoryear{{Ahn} et~al.,}{{Ahn} et~al.}{2014}]{sloan10}
{Ahn} C.~P.,  et~al., 2014, \mn@doi [ApJs] {10.1088/0067-0049/211/2/17}, \href
  {http://adsabs.harvard.edu/abs/2014ApJS..211...17A} {211, 17}

\bibitem[\protect\citeauthoryear{{Akritas} \& {Bershady}}{{Akritas} \&
  {Bershady}}{1996}]{bces}
{Akritas} M.~G.,  {Bershady} M.~A.,  1996, \mn@doi [ApJ] {10.1086/177901},
  \href {http://adsabs.harvard.edu/abs/1996ApJ...470..706A} {470, 706}

\bibitem[\protect\citeauthoryear{{Allen}, {Evrard}  \& {Mantz}}{{Allen}
  et~al.}{2011}]{allen}
{Allen} S.~W.,  {Evrard} A.~E.,   {Mantz} A.~B.,  2011, \mn@doi [ARA&A]
  {10.1146/annurev-astro-081710-102514}, \href
  {http://adsabs.harvard.edu/abs/2011ARA%26A..49..409A} {49, 409}

\bibitem[\protect\citeauthoryear{{Blanton} \& {Roweis}}{{Blanton} \&
  {Roweis}}{2007}]{blanton2}
{Blanton} M.~R.,  {Roweis} S.,  2007, \mn@doi [AJ] {10.1086/510127}, \href
  {http://adsabs.harvard.edu/abs/2007AJ....133..734B} {133, 734}

\bibitem[\protect\citeauthoryear{{Blanton} et~al.,}{{Blanton}
  et~al.}{2003}]{blanton}
{Blanton} M.~R.,  et~al., 2003, \mn@doi [ApJ] {10.1086/375776}, \href
  {http://adsabs.harvard.edu/abs/2003ApJ...592..819B} {592, 819}

\bibitem[\protect\citeauthoryear{{B{\"o}hringer} et~al.,}{{B{\"o}hringer}
  et~al.}{2000}]{noras}
{B{\"o}hringer} H.,  et~al., 2000, \mn@doi [ApJs] {10.1086/313427}, \href
  {http://adsabs.harvard.edu/abs/2000ApJS..129..435B} {129, 435}

\bibitem[\protect\citeauthoryear{{B{\"o}hringer} et~al.,}{{B{\"o}hringer}
  et~al.}{2004}]{reflex}
{B{\"o}hringer} H.,  et~al., 2004, \mn@doi [A\&A] {10.1051/0004-6361:20034484},
  \href {http://adsabs.harvard.edu/abs/2004A%26A...425..367B} {425, 367}

\bibitem[\protect\citeauthoryear{{Cai}, {Angulo}, {Baugh}, {Cole}, {Frenk}  \&
  {Jenkins}}{{Cai} et~al.}{2009}]{orcamock}
{Cai} Y.-C.,  {Angulo} R.~E.,  {Baugh} C.~M.,  {Cole} S.,  {Frenk} C.~S.,
  {Jenkins} A.,  2009, \mn@doi [MNRAS] {10.1111/j.1365-2966.2009.14402.x},
  \href {http://adsabs.harvard.edu/abs/2009MNRAS.395.1185C} {395, 1185}

\bibitem[\protect\citeauthoryear{{Carliles}, {Budav{\'a}ri}, {Heinis}, {Priebe}
   \& {Szalay}}{{Carliles} et~al.}{2010}]{randomforests}
{Carliles} S.,  {Budav{\'a}ri} T.,  {Heinis} S.,  {Priebe} C.,   {Szalay}
  A.~S.,  2010, \mn@doi [ApJ] {10.1088/0004-637X/712/1/511}, \href
  {http://adsabs.harvard.edu/abs/2010ApJ...712..511C} {712, 511}

\bibitem[\protect\citeauthoryear{{Clerc}, {Sadibekova}, {Pierre}, {Pacaud}, {Le
  F{\`e}vre}, {Adami}, {Altieri}  \& {Valtchanov}}{{Clerc}
  et~al.}{2012}]{xclass}
{Clerc} N.,  {Sadibekova} T.,  {Pierre} M.,  {Pacaud} F.,  {Le F{\`e}vre}
  J.-P.,  {Adami} C.,  {Altieri} B.,   {Valtchanov} I.,  2012, \mn@doi [MNRAS]
  {10.1111/j.1365-2966.2012.21153.x}, \href
  {http://adsabs.harvard.edu/abs/2012MNRAS.423.3561C} {423, 3561}

\bibitem[\protect\citeauthoryear{{Gladders}, {Lopez-Cruz}, {Yee}  \&
  {Kodama}}{{Gladders} et~al.}{1998}]{gladdersredevolution}
{Gladders} M.~D.,  {Lopez-Cruz} O.,  {Yee} H.~K.~C.,   {Kodama} T.,  1998,
  \mn@doi [ApJ] {10.1086/305858}, \href
  {http://adsabs.harvard.edu/abs/1998ApJ...501..571G} {501, 571}

\bibitem[\protect\citeauthoryear{{Hansen}, {Sheldon}, {Wechsler}  \&
  {Koester}}{{Hansen} et~al.}{2009}]{hansen2009}
{Hansen} S.~M.,  {Sheldon} E.~S.,  {Wechsler} R.~H.,   {Koester} B.~P.,  2009,
  \mn@doi [ApJ] {10.1088/0004-637X/699/2/1333}, \href
  {http://adsabs.harvard.edu/abs/2009ApJ...699.1333H} {699, 1333}

\bibitem[\protect\citeauthoryear{{Hao} et~al.,}{{Hao}
  et~al.}{2009}]{gmbcgerror}
{Hao} J.,  et~al., 2009, \mn@doi [ApJ] {10.1088/0004-637X/702/1/745}, \href
  {http://adsabs.harvard.edu/abs/2009ApJ...702..745H} {702, 745}

\bibitem[\protect\citeauthoryear{{Hao} et~al.,}{{Hao} et~al.}{2010}]{gmbcg}
{Hao} J.,  et~al., 2010, \mn@doi [ApJs] {10.1088/0067-0049/191/2/254}, \href
  {http://adsabs.harvard.edu/abs/2010ApJS..191..254H} {191, 254}

\bibitem[\protect\citeauthoryear{{Heymans} et~al.,}{{Heymans}
  et~al.}{2012}]{cfhtlens}
{Heymans} C.,  et~al., 2012, \mn@doi [MNRAS]
  {10.1111/j.1365-2966.2012.21952.x}, \href
  {http://adsabs.harvard.edu/abs/2012MNRAS.427..146H} {427, 146}

\bibitem[\protect\citeauthoryear{{High} et~al.,}{{High} et~al.}{2010}]{high}
{High} F.~W.,  et~al., 2010, \mn@doi [ApJ] {10.1088/0004-637X/723/2/1736},
  \href {http://adsabs.harvard.edu/abs/2010ApJ...723.1736H} {723, 1736}

\bibitem[\protect\citeauthoryear{{Jenkins}, {Frenk}, {White}, {Colberg},
  {Cole}, {Evrard}, {Couchman}  \& {Yoshida}}{{Jenkins} et~al.}{2001}]{jenkins}
{Jenkins} A.,  {Frenk} C.~S.,  {White} S.~D.~M.,  {Colberg} J.~M.,  {Cole} S.,
  {Evrard} A.~E.,  {Couchman} H.~M.~P.,   {Yoshida} N.,  2001, \mn@doi [MNRAS]
  {10.1046/j.1365-8711.2001.04029.x}, \href
  {http://adsabs.harvard.edu/abs/2001MNRAS.321..372J} {321, 372}

\bibitem[\protect\citeauthoryear{{Kloster}, {Rines}, {Svoboda}, {Arnold},
  {Welch}, {Finn}  \& {Vikhlinin}}{{Kloster} et~al.}{2011}]{xrayscale}
{Kloster} D.,  {Rines} K.,  {Svoboda} B.~E.,  {Arnold} R.~L.,  {Welch} T.~J.,
  {Finn} R.~A.,   {Vikhlinin} A.,  2011, in American Astronomical Society
  Meeting Abstracts 217. p. 149.12

\bibitem[\protect\citeauthoryear{{Koester} et~al.,}{{Koester}
  et~al.}{2007}]{maxbcg}
{Koester} B.~P.,  et~al., 2007, \mn@doi [ApJ] {10.1086/512092}, \href
  {http://adsabs.harvard.edu/abs/2007ApJ...660..221K} {660, 221}

\bibitem[\protect\citeauthoryear{Kraft}{Kraft}{1988}]{slsqp}
Kraft D.,  1988, A Software Package for Sequential Quadratic Programming.
Deutsche Forschungs- und Versuchsanstalt f{\"u}r Luft- und Raumfahrt K{\"o}ln:
  Forschungsbericht, Wiss. Berichtswesen d. DFVLR

\bibitem[\protect\citeauthoryear{{Lloyd-Davies} et~al.,}{{Lloyd-Davies}
  et~al.}{2011}]{xcsxray}
{Lloyd-Davies} E.~J.,  et~al., 2011, \mn@doi [MNRAS]
  {10.1111/j.1365-2966.2011.19117.x}, \href
  {http://adsabs.harvard.edu/abs/2011MNRAS.418...14L} {418, 14}

\bibitem[\protect\citeauthoryear{{Magnier} et~al.,}{{Magnier}
  et~al.}{2013}]{panstarrs}
{Magnier} E.~A.,  et~al., 2013, \mn@doi [ApJs] {10.1088/0067-0049/205/2/20},
  \href {http://adsabs.harvard.edu/abs/2013ApJS..205...20M} {205, 20}

\bibitem[\protect\citeauthoryear{{Mehrtens} et~al.,}{{Mehrtens}
  et~al.}{2012}]{xcsoptical}
{Mehrtens} N.,  et~al., 2012, \mn@doi [MNRAS]
  {10.1111/j.1365-2966.2012.20931.x}, \href
  {http://adsabs.harvard.edu/abs/2012MNRAS.423.1024M} {423, 1024}

\bibitem[\protect\citeauthoryear{{Miller} et~al.,}{{Miller} et~al.}{2005}]{c4}
{Miller} C.~J.,  et~al., 2005, \mn@doi [AJ] {10.1086/431357}, \href
  {http://adsabs.harvard.edu/abs/2005AJ....130..968M} {130, 968}

\bibitem[\protect\citeauthoryear{{Murphy}, {Geach}  \& {Bower}}{{Murphy}
  et~al.}{2012}]{orca}
{Murphy} D.~N.~A.,  {Geach} J.~E.,   {Bower} R.~G.,  2012, \mn@doi [MNRAS]
  {10.1111/j.1365-2966.2011.19782.x}, \href
  {http://adsabs.harvard.edu/abs/2012MNRAS.420.1861M} {420, 1861}

\bibitem[\protect\citeauthoryear{{Peebles}}{{Peebles}}{1980}]{peebles}
{Peebles} P.~J.~E.,  1980, {The large-scale structure of the universe}.
Princeton University Press

\bibitem[\protect\citeauthoryear{{Planck Collaboration} et~al.,}{{Planck
  Collaboration} et~al.}{2014}]{planckclusters}
{Planck Collaboration} et~al., 2014, \mn@doi [A\&A]
  {10.1051/0004-6361/201321523}, \href
  {http://adsabs.harvard.edu/abs/2014A%26A...571A..29P} {571, A29}

\bibitem[\protect\citeauthoryear{{Reichardt} et~al.,}{{Reichardt}
  et~al.}{2013}]{spts}
{Reichardt} C.~L.,  et~al., 2013, \mn@doi [ApJ] {10.1088/0004-637X/763/2/127},
  \href {http://adsabs.harvard.edu/abs/2013ApJ...763..127R} {763, 127}

\bibitem[\protect\citeauthoryear{{Romer}, {Viana}, {Liddle}  \& {Mann}}{{Romer}
  et~al.}{2001}]{romer}
{Romer} A.~K.,  {Viana} P.~T.~P.,  {Liddle} A.~R.,   {Mann} R.~G.,  2001,
  \mn@doi [ApJ] {10.1086/318382}, \href
  {http://adsabs.harvard.edu/abs/2001ApJ...547..594R} {547, 594}

\bibitem[\protect\citeauthoryear{{Rozo} \& {Rykoff}}{{Rozo} \&
  {Rykoff}}{2014}]{redmapper2}
{Rozo} E.,  {Rykoff} E.~S.,  2014, \mn@doi [\apj] {10.1088/0004-637X/783/2/80},
  \href {http://adsabs.harvard.edu/abs/2014ApJ...783...80R} {783, 80}

\bibitem[\protect\citeauthoryear{{Rozo} et~al.,}{{Rozo}
  et~al.}{2010}]{maxbcgcosmo}
{Rozo} E.,  et~al., 2010, \mn@doi [ApJ] {10.1088/0004-637X/708/1/645}, \href
  {http://adsabs.harvard.edu/abs/2010ApJ...708..645R} {708, 645}

\bibitem[\protect\citeauthoryear{{Rykoff} et~al.,}{{Rykoff}
  et~al.}{2008}]{xrayscale2}
{Rykoff} E.~S.,  et~al., 2008, \mn@doi [\apj] {10.1086/527537}, \href
  {http://adsabs.harvard.edu/abs/2008ApJ...675.1106R} {675, 1106}

\bibitem[\protect\citeauthoryear{{Rykoff} et~al.,}{{Rykoff}
  et~al.}{2014}]{redmapper}
{Rykoff} E.~S.,  et~al., 2014, \mn@doi [ApJ] {10.1088/0004-637X/785/2/104},
  \href {http://adsabs.harvard.edu/abs/2014ApJ...785..104R} {785, 104}

\bibitem[\protect\citeauthoryear{{Shanks} \& {Metcalfe}}{{Shanks} \&
  {Metcalfe}}{2012}]{atlas}
{Shanks} T.,  {Metcalfe} N.,  2012, in Science from the Next Generation Imaging
  and Spectroscopic Surveys.

\bibitem[\protect\citeauthoryear{{Sheth}, {Mo}  \& {Tormen}}{{Sheth}
  et~al.}{2001}]{sheth}
{Sheth} R.~K.,  {Mo} H.~J.,   {Tormen} G.,  2001, \mn@doi [MNRAS]
  {10.1046/j.1365-8711.2001.04006.x}, \href
  {http://adsabs.harvard.edu/abs/2001MNRAS.323....1S} {323, 1}

\bibitem[\protect\citeauthoryear{{Song}, {Mohr}, {Barkhouse}, {Warren}, {Dolag}
   \& {Rude}}{{Song} et~al.}{2012}]{highmock}
{Song} J.,  {Mohr} J.~J.,  {Barkhouse} W.~A.,  {Warren} M.~S.,  {Dolag} K.,
  {Rude} C.,  2012, \mn@doi [ApJ] {10.1088/0004-637X/747/1/58}, \href
  {http://adsabs.harvard.edu/abs/2012ApJ...747...58S} {747, 58}

\bibitem[\protect\citeauthoryear{{Tinker} et~al.,}{{Tinker}
  et~al.}{2012}]{cosmoconst}
{Tinker} J.~L.,  et~al., 2012, \mn@doi [ApJ] {10.1088/0004-637X/745/1/16},
  \href {http://adsabs.harvard.edu/abs/2012ApJ...745...16T} {745, 16}

\bibitem[\protect\citeauthoryear{{Tojeiro}, {Percival}, {Heavens}  \&
  {Jimenez}}{{Tojeiro} et~al.}{2011}]{lrgevolution}
{Tojeiro} R.,  {Percival} W.~J.,  {Heavens} A.~F.,   {Jimenez} R.,  2011,
  \mn@doi [MNRAS] {10.1111/j.1365-2966.2010.18148.x}, \href
  {http://adsabs.harvard.edu/abs/2011MNRAS.413..434T} {413, 434}

\bibitem[\protect\citeauthoryear{{Voit}}{{Voit}}{2005}]{galev}
{Voit} G.~M.,  2005, \mn@doi [Rev. Mod. Phys.] {10.1103/RevModPhys.77.207},
  \href {http://adsabs.harvard.edu/abs/2005RvMP...77..207V} {77, 207}

\makeatother
\end{thebibliography}




\appendix

\section{GMPhoRCC Outputs}
\label{sec:gmphorccoutputs}
A full list of the the flags / SDSS quality marker calculations and outputs

\begin{table*}
  \begin{minipage}{\textwidth}
\centering
  \caption[A list of the GMPhoRCC flags relating to the detection of multi-modal distributions.]{A list of the GMPhoRCC flags relating to the detection of multi-modal distributions. Unfortunately as $\sim 70$ percent of cluster have exhibit multi-modal distributions these are insufficient to identify catastrophic failures.}
  \label{table:flagmulti}
  \begin{tabular*}{\textwidth}{l l l}
    \hline\hline
    Name & Value & Description  \\
    \hline
    
     MULTI\textunderscore BANDS                       &  0x00000000001  &   Multiple bands considered            \\
    
     POOR\textunderscore CMR\textunderscore FIT                   &  0x00000000002  &   Poor CMR fit, $\chi^2 > 5$        \\
    
     INAPPROPRIATE\textunderscore Z                   &  0x00000000004  &   One or more of the redshift estimates not appropriate for RS band \\ 
		
     MULTI\textunderscore INITIAL                     &  0x00000000010  &   Multiple peaks in initial z distribution - relative heights $< 5$    \\
      
     MULTI\textunderscore INITIAL\textunderscore AMBIGUOUS        &  0x00000000020  &   Multiple peaks in initial z distribution - relative heights $< 2$    \\
      
     MULTI\textunderscore INITIAL\textunderscore CLOSE            &  0x00000000040  &   Primary and secondary peak within $0.1$ of each other      \\
                    
     MULTI\textunderscore COLOUR                      &  0x00000000100  &   Multiple peaks in the colour distribution - relative heights $< 5$    \\
      
     MULTI\textunderscore COLOUR\textunderscore AMBIGUOUS         &  0x00000000200  &   Multiple peaks in the colour distribution - relative heights $< 2$    \\
      
     MULTI\textunderscore COLOUR\textunderscore CLOSE             &  0x00000000400  &   Primary and secondary peak within $0.2$ mag of each other     \\
                              
     MULTI\textunderscore ZRS                         &  0x00000001000  &   Multiple peaks in the RS z fit - relative heights $< 5$  \\
      
     MULTI\textunderscore ZRS\textunderscore AMBIGUOUS            &  0x00000002000  &   Multiple peaks in the RS z fit - relative heights $< 2$  \\
      
     MULTI\textunderscore ZRS\textunderscore CLOSE                &  0x00000004000  &   Primary and secondary peak within $0.1$ of each other      \\
    \hline
  \end{tabular*}
  \end{minipage}
\end{table*}

\begin{table*}
  \begin{minipage}{\textwidth}
\centering
  \caption[A list of GMPhoRCC flags indicating issues with the redshift or richness estimates.]{A list of GMPhoRCC flags indicating issues with the redshift or richness estimates which give the strongest indication an estimate may be erroneous. Here $\Delta z_{cp}$ and $\Delta z_{cs}$ are respectively photometric and spectroscopic redshift consistency bounds where, for the SDSS DR10, $\Delta z_{cp} = 0.035$ and $\Delta z_{cs}=0.025$.  }
  \label{table:flagzn}
  \begin{tabular*}{\textwidth}{ l l l }
    \hline\hline
    Name & Value & Description  \\
    \hline
					
     SPARCE\textunderscore INITIAL                    &  0x00000010000  &   $< 5$ Galaxies found in the cluster region for the initial z fit \\
    
     SPARCE\textunderscore COLOUR                     &  0x00000020000  &   $< 5$ Galaxies found in the cluster region for the colour fit   \\
      
     SPARCE\textunderscore ZRS         	              &  0x00000040000  &   $< 5$ Galaxies found in the cluster region for the RS z fit \\
		
		 LOW\textunderscore RICHNESS\textunderscore N200\textunderscore 3         &  0x00000100000  &   Low counting richness recovered, $n_{200-count}<3$    \\
    		
     INCONSISTENT\textunderscore Z\textunderscore PHOT            &  0x00000200000  &   $z_{RS}$ and $z_{BCG-phot}$ are inconsistent with each other, \tiny $|z_{RS}-z_{BCG-phot}| > \Delta z_{cp} $    \\
        
     INCONSISTENT\textunderscore Z\textunderscore SPEC            &  0x00000400000  &   $z_{RS}$ and $z_{BCG-spec}$ are inconsistent with each other, \tiny $|z_{RS}-z_{BCG-spec}| > \Delta z_{cs} $ \\
    
     LOW\textunderscore RICHNESS\textunderscore N200\textunderscore 1         &  0x00001000000  &   Low counting richness recovered, $n_{200-count}<1$     \\
    
     INCONSISTENT\textunderscore Z\textunderscore PHOT\textunderscore 2X      &  0x00002000000  &   $z_{RS}$ and $z_{BCG-phot}$ are inconsistent with each other, \tiny $|z_{RS}-z_{BCG-phot}| > 2\Delta z_{cp} $    \\
        
     INCONSISTENT\textunderscore Z\textunderscore SPEC\textunderscore 2X      &  0x00004000000  &   $z_{RS}$ and $z_{BCG-spec}$ are inconsistent with each other, \tiny $|z_{RS}-z_{BCG-spec}| > 2\Delta z_{cs} $    \\
    
				\hline                    
  \end{tabular*}
  \end{minipage}
\end{table*}

\begin{table*}
  \begin{minipage}{\textwidth}
    \centering
  \caption[A list of GMPhoRCC flags relating to the non-detection of a cluster overdensity.]{A list of GMPhoRCC flags relating to the non-detection of a cluster overdensity.}
  \label{table:flagnodectection}
  \begin{tabular*}{\textwidth}{l l l }
    \hline\hline
    Name & Value & Description  \\
    \hline
		                   
    CLUSTER\textunderscore INSIDE\textunderscore MASK\textunderscore 0\textunderscore 5\textunderscore MPC    &  0x00010000000  &   Empty apertures found inside $r < 0.5$h$^{-1}$ Mpc of cluster centre      \\
      
    CLUSTER\textunderscore INSIDE\textunderscore MASK\textunderscore R200             &  0x00020000000  &   Empty apertures found inside $r < r_{200}$  of cluster centre      \\
    
    CLUSTER\textunderscore INSIDE\textunderscore MASK\textunderscore 5\textunderscore AM          &  0x00040000000  &   Empty apertures found inside $r < 5'$ of cluster centre      \\
		
    NO\textunderscore OVERDENSITY\textunderscore INITIAL                  &  0x00100000000  &   No overdensity found in the cluster region for the initial z fit  \\
    
		NO\textunderscore OVERDENSITY\textunderscore COLOUR                   &  0x00200000000  &   No overdensity found in the cluster region for the colour fit    \\
       
    NO\textunderscore OVERDENSITY\textunderscore ZRS                      &  0x00400000000  &   No overdensity found in the cluster region for the RS z fit  \\
      
    NO\textunderscore CLUSTER\textunderscore INITIAL                      &  0x01000000000  &   $0$ Galaxies found in the cluster region for the initial z fit  \\
     
		NO\textunderscore CLUSTER\textunderscore COLOUR                       &  0x02000000000  &   $0$ Galaxies found in the cluster region for the colour fit    \\
            
    NO\textunderscore CLUSTER\textunderscore ZRS                          &  0x04000000000  &   $0$ Galaxies found in the cluster region for the RS z fit  \\
        
    NO\textunderscore DETECTION\textunderscore REDSHIFT                   &  0x10000000000  &   No detection in redshift module         \\
      
    NO\textunderscore DETECTION\textunderscore RICHNESS\textunderscore NGALS          &  0x20000000000  &   No detection in richness, $n_{gals} < 0$ for both counting and luminosity   \\
       
    NO\textunderscore DETECTION\textunderscore RICHNESS\textunderscore N200           &  0x40000000000  &   No detection in richness, $n_{200} < 0$ for both counting and luminosity   \\
      
    NO\textunderscore COVERAGE                                &  0x80000000000  &   No optical coverage   \\
    \hline
  \end{tabular*}
\end{minipage}
\end{table*}

\begin{table*}
  \centering
  \begin{minipage}{\textwidth}
  \caption[A list of the quality markers assigned to clusters based on the GMPhoRCC flags.]{A list of the quality markers assigned to clusters based on the GMPhoRCC flags.}
  \label{table:markers1}
  \begin{tabular}{c l l}
    \hline\hline
    Quality &   flags Value & Description \\
   \hline
    $-1$  &  0x80000000000 $\leq$ flags                   & no optical coverage \\
   
    $0$  &  0x01000000000 $\leq$ flags $<$ 0x80000000000  & no characterisation found \\
    
    $1$  &  0x00001000000 $\leq$ flags $<$ 0x01000000000  & $n_{200}<1$, large redshift inconsistencies, field masking issues  \\
    
    $2$  &  0x00000100000 $\leq$ flags $<$ 0x00001000000  & $n_{200}<3$, small redshift inconstancies   \\
   
    $3$  &  \hspace{22.68mm}     flags $<$ 0x00000100000  & clean \\
    \hline
  \end{tabular}
\end{minipage}
\end{table*}

\begin{table*}
  \centering
  \caption[A list of the subsets of clusters based on the GMPhoRCC quality marker.]{A list cluster subsets based on the GMPhoRCC quality markers used remove potentially erroneous characterisations.}
  \label{table:qualsub}
  \begin{tabular*}{\textwidth}{p{58pt} p{36pt} p{374pt} }
    \hline\hline
    Subset      &   Quality   & Description\\
    \hline
    Detection   & $\geq1$ &  All clusters considered to have been detected \newline i.e. estimates were found for both redshift and richness \\
    
    Mid         & $\geq2$ &  An intermediate subset removing the worst outliers \newline i.e. removing clusters with very low richness or large discrepancies between redshift estimates \\
    
    Clean       & $\geq3$ &  The cleanest subset removing the majority of outliers \newline i.e. removing cluster with low richness and discrepancies between redshift estimates \\
    \hline
  \end{tabular*}
\end{table*}

\begin{table*}
 \begin{minipage}{\textwidth}
   \centering
  \caption[A list of outputs generated by the GMPhoRCC.]{A list of outputs generated by GMPhoRCC using SDSS DR10 photometry. These properties are given for the primary and secondary cluster with the latter denoted by a `\textunderscore sec' suffix. In the case where GMPhoRCC was unable to determine a property a default value of $-1$ is used. While the redshift labels are specific to SDSS DR10 these can be adjusted to match any optical input.}
  \label{table:output}
  
    \begin{tabular*}{\textwidth}{p{90pt} p{390pt}}
      \hline\hline
      Name & Description   \\
     
      \hline
       band	                          &  The red sequence colour used to detect the cluster. $0 = g-r$, $1 = r-i$, $2 = i-z$. \\
      
       size                               &  The angular radius in arc minutes of the initial aperture used to model the red sequence. \\
      
       z\textunderscore initial                       &  The position of the peak of the initial redshift distribution.   \\
      
       z\textunderscore initial\textunderscore peak               &  The size of the peak of the initial redshift distribution (galaxies . arc minutes$^{-2}$). \\
      
       z\textunderscore initial\textunderscore errorm(p)          &  $1$-sigma error on the peak in the `minus' (`positive') direction. \\
      
       z\textunderscore initial\textunderscore info               &  A flag based on how the error was determined. 0 = No issues, 1 = Extrapolation needed due to multiple peaks. \\
      
       rs\textunderscore colour\textunderscore (peak,error,info)  &  The position, amplitude and error of the peak in the initial red sequence colour distribution. \\
      
       z\textunderscore rs\textunderscore (peak,error,info)       &  The position, amplitude and error of the peak in the red sequence photometric redshift distribution. \\
      
       BCG\textunderscore objID                       &  The objID of the BCG.  \\
      
       BCG\textunderscore dis                         &  The angular distance in arc minutes of the BCG from the cluster centre. \\
      
       z\textunderscore BCG\textunderscore best\textunderscore (err)          &  The best redshift with error of the BCG, spectra if available, photometric otherwise. \\
      
       z\textunderscore BCG\textunderscore phot\textunderscore (err)          &  The photometric redshift of the BCG    \\
      
       z\textunderscore BCG\textunderscore spec\textunderscore (err)          &  The spectroscopic redshift of the BCG.   \\
      
       z\textunderscore gals\textunderscore spec\textunderscore (err)         &  A spectroscopic cluster redshift based on the spectra of the $5$ brightest galaxies on the red sequence. \\
      
       z\textunderscore gals\textunderscore spec\textunderscore no            &  The number of galaxies available with spectra.  \\
      
       cmr\textunderscore grad\textunderscore (err)               &  The gradient of the red sequence CMR. \\
      
       cmr\textunderscore intercept\textunderscore (err)          &  The intercept of the red sequence CMR.  \\
      
       cmr\textunderscore width                       &  The intrinsic width the red sequence CMR.  \\
      
       ngals\textunderscore count\textunderscore (err)            &  $n_{gals-count}$, the background-subtracted number of galaxies inside $0.5$h$^{-1}$ Mpc on the red sequence with poissonian error. \\
      
       ngals\textunderscore lum\textunderscore (err)              &  $n_{gals-lum}$, the background-subtracted richness inside $0.5$h$^{-1}$ Mpc from integrating a LF with poissonian error. \\
      
       r200\textunderscore mpch-1                     &  $r_{200}$ in h$^{-1}$ Mpc.   \\
      
       n200\textunderscore count\textunderscore (err)             &  $n_{200-count}$, the background-subtracted number of galaxies inside $r_{200}$ on the red sequence with error. \\
      
       n200\textunderscore lum\textunderscore (err)               &  $n_{200-lum}$, the background-subtracted richness inside $r_{200}$ from integrating a LF with error. \\
      
       flags                              &  A hexadecimal combination of the GMPhoRCC flags.   \\
      
       $q$                                &  The quality marker based on the GMPhoRCC flags.\\

      \hline
    \end{tabular*}
  
\end{minipage}
\end{table*}


\bsp	
\label{lastpage}
\end{document}